\documentclass[fleqn,usenatbib]{mnras}

\usepackage{newtxtext,newtxmath}

\usepackage[T1]{fontenc}

\DeclareRobustCommand{\VAN}[3]{#2}
\let\VANthebibliography\thebibliography
\def\thebibliography{\DeclareRobustCommand{\VAN}[3]{##3}\VANthebibliography}

\usepackage{graphicx}	
\usepackage{subcaption}
\usepackage{enumitem}
\usepackage{xspace} 
\usepackage[dvipsnames]{xcolor} 
\usepackage{booktabs}
\usepackage{orcidlink}

\newcommand{\pder}[2]{\frac{\partial#1}{\partial#2}}

\newcommand{\BV}{Brunt-V\"{a}is\"{a}l\"{a}\xspace}
\setcitestyle{citesep={,}}

\title[Impact of Si/O interface on CCSN explosions]{Quantifying the Impact of the Si/O Interface in CCSN Explosions Using the Force Explosion Condition}

\author[L. Boccioli et al.]{
Luca Boccioli,$^{1}$\thanks{E-mail: lbocciol@berkeley.edu}\orcidlink{0000-0002-4819-310X}
Mariam Gogilashvili,$^{2}$ \orcidlink{0000-0002-6944-8052}
Jeremiah Murphy$^{3}$ \orcidlink{0000-0003-1599-5656}
and Evan P. O'Connor$^{4}$ \orcidlink{0000-0002-8228-796X}
\\
$^{1}$Department of Physics, University of California, Berkeley, CA 94720, USA\\
$^{2}$Niels Bohr International Academy and DARK, Niels Bohr Institute, University of Copenhagen, Blegdamsvej 17, 2100, Copenhagen, Denmark\\
$^{3}$Department of Physics, Florida State University, 77 Chieftan Way, Tallahassee, 32306, FL, USA\\
$^{4}$The Oskar Klein Centre, Department of Astronomy, Stockholm University, AlbaNova, SE-106 91 Stockholm, Sweden
}

\date{Accepted XXX. Received YYY; in original form ZZZ}

\pubyear{2025}

\begin{document}
\label{firstpage}
\pagerange{\pageref{firstpage}--\pageref{lastpage}}
\maketitle

\begin{abstract}
The explosion mechanism of a core-collapse supernova is a complex interplay between neutrino heating and cooling (including the effects of neutrino-driven convection), the gravitational potential, and the ram pressure of the infalling material. To analyze the post-bounce phase of a supernova, one can use the generalized Force Explosion Condition (FEC+), which succinctly formalizes the interplay among these four phenomena in an analytical condition, consistent with realistic simulations. In this paper, we use the FEC+ to study the post-bounce phase of 341 spherically symmetric simulations, where convection is included through a time-dependent mixing length approach. We find that the accretion of the Si/O interface through the expanding shock can significantly change the outcome of the supernova by driving the FEC+ above the explosion threshold. We systematically explore this by (i) artificially smoothing the pre-supernova density profile, and (ii) artificially varying the mixing length. In both cases, we find that large-enough density contrasts at the Si/O interface lead to successful shock revival only if the FEC+ is already close to the explosion threshold. 
Furthermore, we find that the accretion of the Si/O interface has a substantial effect on the critical condition for supernova explosions, contributing between 5\% and 15\%, depending on how pronounced the density contrast at the interface is. Earlier studies showed that convection affects the critical condition by 25--30\%, which demonstrates that the accretion of the Si/O interface through the shock can play a nearly comparable role in influencing shock dynamics.
\end{abstract}




\section{Introduction}
The explosion mechanism of a core-collapse supernova (CCSN) has been the topic of many decades of research \citep{Colgate_White1966,Bethe_Wilson1985,Burrows1993_Theory_SN_expl,Herant1994_first2D,Fryer2002_first3D,Janka2012_review_CCSNe,Muller2016_review,Burrows2021_SN_review}. Theoretical, observational, and computational efforts have shed light onto this complex phenomenon, and the theoretical understanding of the explosion of a CCSN has drastically improved in the last decade, thanks to the rapid improvement of supercomputers and, therefore, of detailed simulations \citep{Lentz2015_3D,OConnor2018_3Dprogenitors,Burrows2020_3DFornax,Fernandez2015_3D_SASI,Takiwaki2016_3DnSNe_3D_explosions,Nakamura2024_LS220_3D_suite,Cabezon2018_3Dcomparison,Glas2019_comparison_nu_transport_AA,Summa2018_rot3D_crit_lum,Bollig2021_3D_7minutes}.

Nonetheless, the detailed mechanism of the CCSN explosion is still a matter of active research. Understanding the physical phenomena that lead to an explosion has important consequences in a variety of astrophysical environments. Knowing which stars explode and which ones fail determines the distribution of compact objects in the Universe \citep{Boccioli2024_remnant,Fryer2012_remnant_popsynth,Patton2020_popsynth_prescription,Fryer2022_nu_conv_remnant_masses}. It also determines the thermodynamic conditions that lead to the nucleosynthesis of heavy elements, which in turn determines the chemical enrichment of the interstellar medium \citep{Diehl2021_26Al_Fe60_review,Woosley2002_KEPLER_models}. Finally, understanding the physical phenomena responsible for triggering the explosion can unveil what are the pre-collapse features that are important for the explosion or the failed shock revival \citep{Boccioli2023_explodability}. 

The first breakthrough in studying the CCSN explosion mechanism was made by \citet{Colgate_White1966}, who suggested that neutrinos were responsible for carrying energy from the deep interior of the central proto-neutron star (PNS) to the vicinity of the shock. Later, \citet{Bethe_Wilson1985} refined this picture by introducing the so-called "delayed neutrino mechanism". In their simulations, neutrinos take a few hundred milliseconds after the bounce to transfer enough energy to revive the shock. During this period, the shock stalls at a nearly constant radius. However, with more advanced models for describing the state of matter at supranuclear densities and interactions between neutrinos and matter \citep{Bruenn1985,Lattimer1991_LS}, the explosion could not be obtained in spherical symmetry anymore. The reason is the lack of multi-dimensional effects, and state-of-the-art multidimensional simulations indeed show self-consistent explosions \citep{Lentz2015_3D,Couch2015_3D_final_stages,Burrows2020_3DFornax,Nakamura2024_LS220_3D_suite,Summa2018_rot3D_crit_lum,Bollig2021_3D_7minutes}. In particular, in the last two decades neutrino-driven convection has been shown to play a crucial role for the explosion dynamics \citep{Foglizzo2006_SASI,Murphy2013_turb_in_CCSNe,Radice2016,Radice2018_turbulence,Abdikamalov2015_turb_SASI_3D}. The importance of this effect motivated the development of parametric models to include neutrino driven convection and turbulent dissipation effects in spherically symmetric, 1D simulations. These parametric models are known as 1D+ simulations and over the years, they were able to produce successful explosions \citep{Mabanta2018_MLT_turb,Couch2020_STIR,Boccioli2021_STIR_GR,Sasaki2024_STIR_diffusion}.

The advantage of such models is that they are computationally very affordable, and therefore allow for systematic studies of the explosion mechanism itself. One of the first attempts at deriving an explosion condition was the seminal work of \citet{Burrows1993_Theory_SN_expl}, who derived a critical luminosity condition based on the idea that, given a mass accretion rate, there is a maximum neutrino luminosity above which a stalled shock solution is not realizable, and therefore an explosion can develop. Similar studies were later carried out over the years \citep{Murphy2008_crit_lum_2D,Yamasaki2005_rot_crti_cond,Janka2012_review_CCSNe,Pejcha2012_antesonic_condition,Summa2016_prog_dependence_vertex,Summa2018_rot3D_crit_lum,Murphy2017_int_cond,Gogilashvili2022_FEC}, oftentimes guided by the results of multi-dimensional simulations.

Recently, a few studies \citep{Boccioli2023_explodability,Wang2022_prog_study_ram_pressure} have pointed out that the magnitude of the density drop located near the Si/O interface can be used to predict the outcome of the explosion. Even before these studies, the idea that the accretion of the Si/O interface onto the shock could facilitate the explosion was well established (references). Because of the sudden change in composition, there is usually a large density drop at the Si/O interface that, when accreted through the shock, leads to a sudden drop in ram pressure, facilitating a rapid shock expansion that, under the right conditions, can turn into a runaway explosion.

In this paper, we will analyze how the accretion onto the shock of the density drop, which typically occurs near the Si/O interface, affects the explosion. In particular, this manuscript is structured as follows: in Section~\ref{sec:FEC} we describe the FEC+, in Section~\ref{sec:SiO_accr} we discuss the accretion of the Si/O interface through the shock, and in Section~\ref{sec:numerics} we describe our numerical setup. Then, in Section~\ref{sec:FEC_many_progs} we analyze the explosion of 341 1D+ simulations, and in Section~\ref{sec:interplay} we study how the strength of neutrino-driven convection and the magnitude of the density drop at the Si/O interface change the FEC+ and therefore the explosion. Finally, we discuss the results in Section~\ref{sec:conclusions}.

\section{Methods}
\subsection{The generalized Force Explosion Condition}
\label{sec:FEC}
Over the years many have suggested explosion criteria for successful CCSN explosions \citep{Ertl2016_explodability,Pejcha2012_antesonic_condition,Muller2016_prog_connection,Wang2022_prog_study_ram_pressure,Boccioli2023_explodability}. Over three decades ago, \citet{Burrows1993_Theory_SN_expl} pointed out that the stalled shock phase is a steady-state, boundary-value problem. They parameterized this problem by the neutrino luminosity, $L_\nu$, and mass accretion rate, $\dot{M}$, and they found a critical curve in these parameters, below which stalled shock solutions exist and above this critical curve there are no steady-state solutions.  They suggested that the solutions above this critical curve are most likely explosive. 

Inspired by the insight of \citet{Burrows1993_Theory_SN_expl}, \citet{Murphy2017_int_cond} utilized semi-analytic techniques to propose an integral condition for supernova explosions. This work suggests that the dimensional integral of the momentum equation, denoted as $\Psi$, is helpful for determining the existence of stalled-shock solutions. The relationship between the shock velocity and the parameter $\Psi$ indicates that a stalled solution exists if $\Psi = 0$. This force explosion condition is consistent with critical neutrino luminosity conditions and one-dimensional simulations and implies that more factors than just neutrino luminosity and mass accretion rate influence explosion dynamics. They show that the integral explosion condition corresponds to a critical hypersurface where the physical dimensions are $L_{\nu}$, $M_{\rm NS}$, $R_{\rm NS}$, and $\dot{M}$.

\citet{Gogilashvili2022_FEC} derived an analytic model for the force explosion condition, beginning with fundamental hydrodynamic equations and identifying that the explosion condition depends on two dimensionless parameters instead of four dimensional parameters. The force explosion condition (FEC) is expressed as:
\begin{equation}
    \label{eq:FEC}
    \tilde{L}_\nu \tau_g - a \tilde{\kappa} \geq b,
\end{equation} 
where $\tilde{L}_\nu \tau_g = L_{\nu} \tau_g R_{\rm NS}/ ( G \dot{M} M_{\rm NS})$ represents the net neutrino power normalized by accretion power, and $\tilde{\kappa} = \kappa \dot{M} / \sqrt{G M_{\rm NS} R_{\rm NS}}$ is a dimensionless neutrino opacity. The coefficients $a$ and $b$ mostly depend upon the density profile of the material behind the shock and can be estimated analytically to be $a \sim 0.05$ and $b \sim 0.51$. It is crucial to emphasize that the theory provides an analytic functional form of the FEC. While the coefficients can be estimated analytically, these estimates might not be accurate due to the approximations and assumptions of the model and thus might need proper calibration with CCSN simulations. 

Notice that in this work we use the net neutrino heating rate in the gain region per unit time $\tilde{\dot{Q}}_\nu$ in place of the first term in Eq.~\eqref{eq:FEC} $\tilde{L}_\nu \tau_g$, as done in \citet{Gogilashvili2023_FEC_GR1D} when using self-consistent simulations to calculate the FEC. The reason behind this is that in the original semi-analytical FEC formulation, the term $L_\nu \tau_g$ is meant to represent the net neutrino heating rate in the gain region $\dot{Q}_\nu$, which can however be directly calculated from our simulations.
 
The validity of the FEC has been checked through three approaches. \citet{Gogilashvili2022_FEC} first, compared the FEC with the semi-analytic condition of \citet{Burrows1993_Theory_SN_expl}, leading to a numerical fit for $a = 0.06$ and $b = 0.38$, which closely matched their analytical estimates. Second, they demonstrated that the FEC accurately predicts explosion conditions in one-dimensional light-bulb simulations. In follow-up research, \citet{Gogilashvili2023_FEC_GR1D} adapted the FEC for simulations using actual neutrino transport, such as those conducted with the \texttt{GR1D} code \citep{OConnor2010, OConnor2015}. Their tests confirmed that the FEC reliably predicts explosion conditions in spherically symmetric supernova simulations, highlighting its robustness as a diagnostic tool.

Motivated by the accuracy of the spherical force explosion condition, \citet{Gogilashvili2023_FEC+} generalized the FEC to incorporate multi-dimensional effects such as neutrino-driven convection and turbulent dissipation in a simple model. This generalized condition, termed FEC+, is expressed as 
\begin{equation}
    \label{eq:FEC+}
    \tilde{\dot{Q}}_\nu + \tilde{W}_b - a \tilde{\kappa} + c \left \langle \tilde{R}^r_r \right \rangle = b.
\end{equation}

It depends on two additional dimensionless parameters: $\tilde{W}_b = W_b R_{\rm NS} / (G \dot{M} M_{\rm NS})$, representing the dimensionless buoyant driving, and $\tilde{R}^r_r = R^r_r R_{\rm NS} / (G M_{\rm NS})$, representing the dimensionless Reynolds stress. These additional terms (mostly buoyant driving) reduce the net neutrino heating required for explosion by $26\%$, a reduction consistent with published studies \citep{Murphy2011, Murphy2013_turb_in_CCSNe, Mabanta2018_MLT_turb}. This finding highlights the significance of multi-dimensional effects in supernova explosion dynamics. Consequently, the FEC+ offers a more comprehensive and accurate framework for describing the explodability of multi-dimensional simulations, enhancing our understanding of the underlying physical processes. 

As mentioned above, in the simple spherically symmetric case the coefficients in Eq. \eqref{eq:FEC} can be fitted using a simple steady-state model, as derived by \citep{Gogilashvili2022_FEC}. However, for the more complicated model that includes convection (i.e. the FEC+ in eq. \eqref{eq:FEC+}), a more accurate calibration is needed, and the same holds for applying the FEC+ to multi-dimensional simulations. For the present paper, we therefore explicitly calculate the coefficients $a$ and $c$ in Eq. \eqref{eq:FEC+} based on the expressions found in Appendix A of \citet{Gogilashvili2023_FEC+}. As a consequence, there is as yet no \textit{a priori} derivation of the threshold value, i.e. the coefficient $b$.  Instead, we empirically determine the threshold $b$ by analyzing hundreds of 1D+ simulations in the next section. In practice, the threshold is found to be around 0.28--0.3.

\subsection{The role of the Si/O interface accretion}
\label{sec:SiO_accr}

The presence of density discontinuities in the pre-SN progenitor can play a significant role in the explosion \citep{Ertl2016_explodability,Wang2022_prog_study_ram_pressure, Boccioli2023_explodability,Tsang2022_ML_explodability}. In particular, the density discontinuity located near the Si/O interface is responsible for triggering the explosion in the vast majority of progenitors. However, as we will quantitatively show in the remainder of this paper, the accretion of this interface onto the shock triggers the explosion only if the star is already close to exploding (in other words, the FEC+ is close to the threshold). This requires multi-dimensional effects and is therefore only possible in 2D, 3D, or 1D+ simulations. Specifically, \cite{Wang2022_prog_study_ram_pressure} and \citet{Boccioli2023_explodability} found that one can build a criterion based on the density drop at the Si/O interface that can successfully predict the explodability of $> 90\%$ progenitors. In particular, both studies independently showed that when the density drop is such that $\delta \rho_{\rm Si/O}^2 / \rho_{\rm Si/O}^2 > 0.08$ \citep{Boccioli2023_explodability} (or, equivalently, $\delta P_{\rm ram} / P_{\rm ram} > 0.28$ \citep{Wang2022_prog_study_ram_pressure}), an explosion will ensue. However, for high compactness progenitors, which are more common at low metallicities, \cite{Boccioli2024_remnant} showed that, even in the absence of this density drop, a successful shock revival would still occur, hinting that the explosion mechanism for these high-compactness stars is slightly different. 

The Si/O interface usually occurs in correspondence with a pocket of oxygen inside the silicon-sulfur shell, which is why it should more precisely be referred to as a Si/O interface. For the sake of brevity, we will however refer to it simply as a Si/O interface in the remainder of this paper. Usually, the shock reaches this interface a few hundred milliseconds (typically 100-400 ms) after bounce, i.e. during the stalled shock phase. The sudden drop in density causes the ram pressure of the infalling material to suddenly decrease, and the shock is therefore able to expand to larger radii. Under favorable conditions, this expansion turns into a runaway explosion, causing a successful shock revival. This has been shown by several multi-dimensional simulations in the past few years \citep{Lentz2015_3D,Muller2020_SNReview,Nakamura2022_3DnSNe_binary_star_1987a,Andresen2017_GW_Garching,Burrows2020_3DFornax,Summa2016_prog_dependence_vertex,Nakamura2024_LS220_3D_suite}, where a successful shock revival occurs right after the shock reaches the Si/O interface. Notice that, especially for 2D simulations, there are instances where the Si/O interface is accreted at times between $200-400$~ms after bounce and the explosion instead is triggered much later, at times between $700$~ms and 1~s \citep{Marek2009_SASI_diag_ene,OConnor2018_2D_M1}. Moreover, both in 2D and 3D simulations, different nuclear equations of state can lead to earlier or later explosions, depending on the details of the microphysics \citep{Eggenberger2021_EOS_dependence_GW_2D,Janka2022_EOS_and_dynamics}.

This suggests that, after the interface is accreted onto the shock, if the change in ram pressure is large enough to disrupt the quasi-steady state, then an explosion will occur. Otherwise, the perturbation might be large enough to move the standing accretion shock to larger radii, but not large enough to completely disrupt the quasi-steady state. In that case, the shock will stall at larger radii and, eventually, it will slowly recede and the explosion will fail (with a few exceptions, as discussed in Section~\ref{sec:FEC+_subset}). In the remainder of this paper, we will quantitatively test this hypothesis using the generalized force explosion condition (hereafter FEC+) described in Section~\ref{sec:FEC}. 

\subsection{Numerical setup}
\label{sec:numerics}
To analyze how the accretion of the Si/O interface affects the explosion, we employed a series of simulations performed with the open-source code \texttt{GR1D} \citep{OConnor2010,OConnor2015}, modified as described in \cite{Boccioli2021_STIR_GR} to include the effect of neutrino driven convection via the Reynolds decomposition model based on time-dependent mixing-length theory STIR \citep{Couch2020_STIR}. The equation of state for all simulations is the SFHo equation of state for nuclear matter \citep{Hempel2010_HS_RMF,Steiner2013_SFHo}, and neutrino opacities from \citet{Bruenn1985}, with weak-magnetism and recoil corrections from \citet{Horowitz2002}, and the virial correction for neutrino-nucleon scattering from \citet{Horowitz2017_virial}. Neutrino-electron scattering opacities follow the inelastic treatment of \citep{Bruenn1985}. The neutrino transport solves the two-moment equations and uses the analytic M1 closure scheme to relate the radiation pressure to the radiation energy density and momentum \citep{OConnor2015}. The neutrino spectrum is resolved with 18 energy groups logarithmically spaced from 1 to 280 MeV.

The spatial resolution was the same for all of the simulated progenitors (with the only exception of $<10$ progenitors of very low compactness, with $M_{\rm ZAMS} < 11$). We adopted a grid of 700 zones linearly spaced up to 20 km, with a resolution of 300 m, and then logarithmically spaced out to 15,000 km. The decision to use the same spatial grid for all progenitors was motivated by the fact that the extra heating from STIR has a weak dependence on the spatial resolution, as shown in \citet{Boccioli2022_EOS_effect}, particularly behind the shock.

\subsection{The STIR model}
The STIR model \citep{Couch2020_STIR} is based on a Reynolds decomposition of the Euler equations, and it relies on a mixing-length-like closure that relates higher-order turbulent correlations to lower-order turbulent correlations. The STIR model is a local algebraic model for which the equations are closed using algebraic relationships at a local level. Other models explored different closures \citep{Murphy2013_turb_in_CCSNe,Mabanta2019_turbulence_in_CCSN,Muller2019_STIR}.

The main modification that STIR introduces is to the internal energy equation, although extra diffusive terms must also be added to the equations describing the evolution of the electron fraction and the neutrino energy. STIR evolves the internal energy and the turbulent energy separately, and therefore the evolution equation of the total energy is the combination of those two (i.e. equations 26 and 29 from \cite{Couch2020_STIR}):
\begin{equation}
\begin{split}
\label{eq:total_energy}
\pder{(\rho e_{\rm tot})}{ t} &+ \frac{1}{r^2}\pder{}{r} [r^2 v_r(\rho e_{\rm tot} + P + P_{\rm turb}) - r^2\rho D \nabla e_{\rm tot}] \\ 
&= -\rho v_r g + Q_\nu -\rho v_{\rm turb}^2 \pder{v_r}{r} + \rho v_{\rm turb} \omega_{\rm BV}^2 \Lambda_{\rm mix}.
\end{split}
\end{equation}
where $e_{\rm tot} = e + v_{\rm turb}^2$, and $P_{\rm turb} = \rho v_{\rm turb}^2$. In the above equations, $e$ includes the contributions from both internal and kinetic energy. Notice that \texttt{GR1D} solves the general relativistic version of the above equation, but we show the Newtonian version for simplicity. The quantities $\omega_{\rm BV}$ and $\Lambda_{\rm mix}$ are the \BV frequency and the mixing length, respectively, defined as:
\begin{align}
    \label{eq:BV_eff}
    \omega^2_{\rm BV} &= \left(g - v \pder{v}{r}\right)\left(\frac{1}{\rho} \pder{\rho(1+\epsilon)}{r} - \frac{ 1}{\rho c_s^2} \pder{P}{r} \right), \\
    \label{eq:lambda_mix}
    \Lambda_{\rm mix} &= \alpha_{\rm MLT} \frac{P}{\rho g},
\end{align}
where we again have not included most of the general relativistic corrections, for simplicity. However, it should be highlighted that one should use the total energy $\rho(1+\epsilon)$ in the expression of the \BV frequency, as shown in Eq. \eqref{eq:BV_eff}, rather than simply $\rho$. This can change the magnitude of the \BV by more than 20\%. Equations \eqref{eq:BV_eff} and \eqref{eq:lambda_mix} also show the dependence of STIR on the main parameter of the model: $\alpha_{\rm MLT}$. A larger value of $\alpha_{\rm MLT}$ increases the magnitude of the mixing length, which as can be seen from Eq. \eqref{eq:total_energy} corresponds to introducing a larger source of turbulent energy, and therefore stronger convection and subsequent explosion.

The extra terms due to STIR are the last two terms on the RHS of eq. \ref{eq:total_energy}, and the last two terms on the LHS. The terms on the LHS, i.e. the diffusive terms, have a smaller (although non-negligible) effect on the overall dynamics compared to the terms on the RHS \citep{Sasaki2024_STIR_diffusion}. If one applies the FEC+ to this model \citep{Gogilashvili2023_FEC+}, then one finds that $\tilde{W}_b$ in eq. \eqref{eq:FEC+} corresponds to the last term on the RHS of eq. \eqref{eq:total_energy}, whereas $\langle \tilde{R}^r_r \rangle$ corresponds to the advection of $P_{\rm turb}$ in eq. \eqref{eq:total_energy}. Notice that the shear term (i.e. the third term on the RHS of eq. \eqref{eq:total_energy}), is quite small, and therefore ignored in the expression of the FEC+.

\begin{figure*}
\centering
\begin{subfigure}{0.48\textwidth}
    \includegraphics[width=\textwidth]{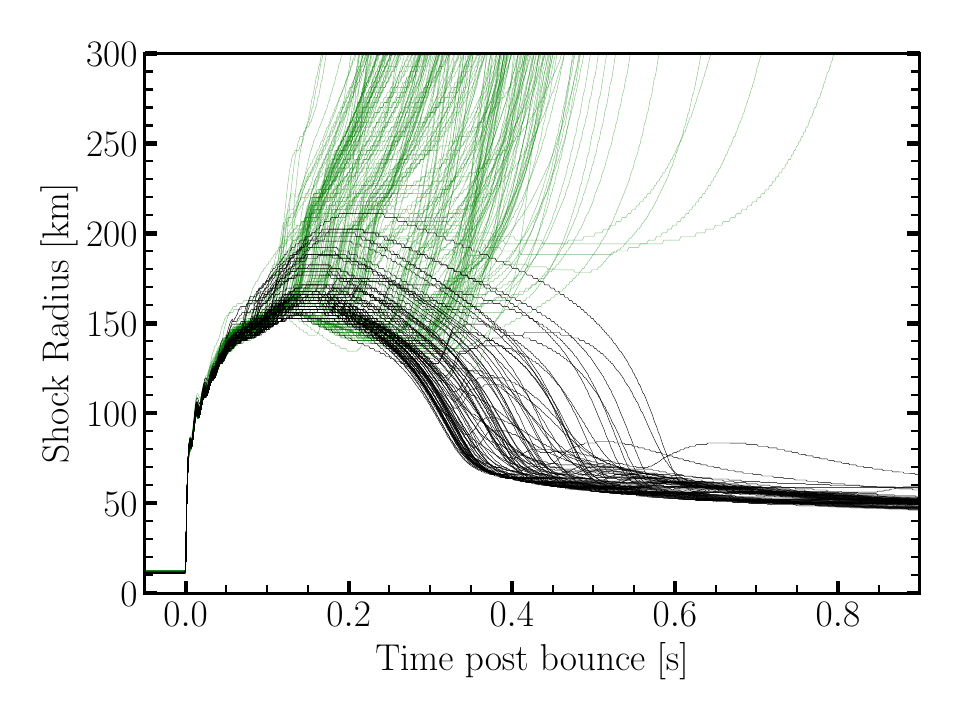}
    \caption{}
\end{subfigure}
\hfill
\begin{subfigure}{0.48\textwidth}
    \includegraphics[width=\textwidth]{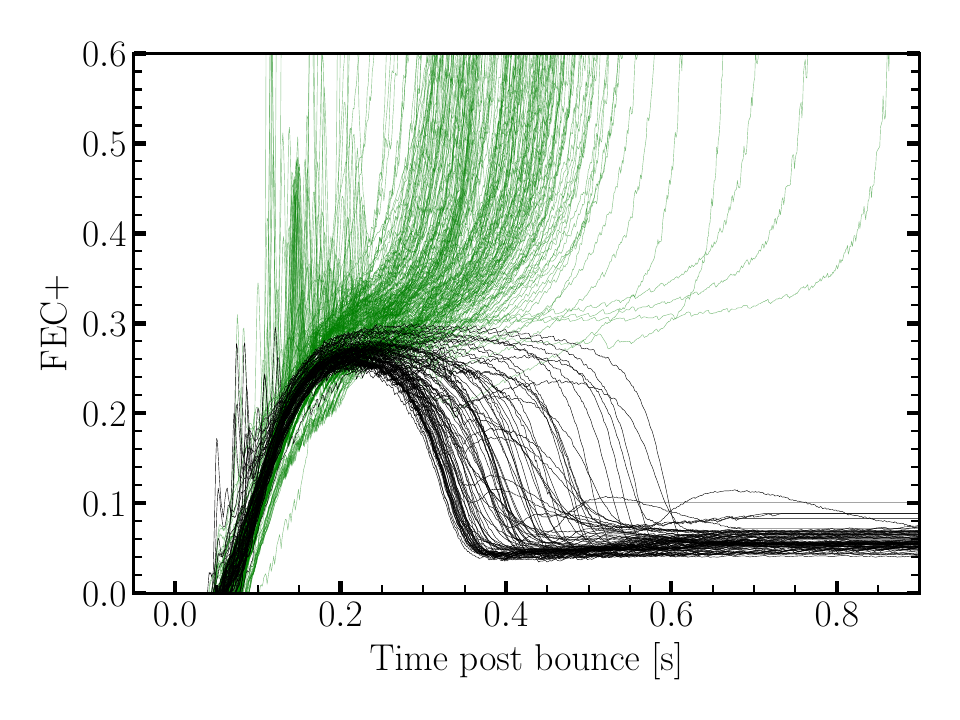}
    \caption{}
\end{subfigure}
        
\caption{This Figure shows the shock evolution (left panel) and FEC+ (right panel) of the collapse and subsequent post-bounce evolution computed by \citep{Boccioli2024_remnant} of 341 progenitors at three different metallicities. Green lines indicate progenitors with successful shock revival, whereas black lines indicate progenitors for which the explosion has not been launched. There is a clear bifurcation in FEC+, and empirically it can be seen that progenitors who reach a threshold around $\sim 0.28\text{--}0.3$ explode. Those that do not reach this threshold do not explode.}
\label{fig:all_progs}
\end{figure*}

\section{Explosion dynamics of 341 progenitors}
\label{sec:FEC_many_progs}
One of the primary aims of this manuscript is to use FEC+ to understand why the Si/O interface facilitates an explosion for most 1D+ STIR simulations and why the Si/O interface does not trigger explosions for other progenitors. The simulations for this study are described in  \cite{Boccioli2024_remnant}. In particular, Figure~\ref{fig:all_progs} shows the evolution of the shock radius and the FEC+ for 341 \texttt{GR1D+} simulations of KEPLER progenitors from \citet{Sukhbold2016_explodability} and \citet{Woosley2002_KEPLER_models}, with masses ranging from 9 $M_\odot$ to 120 M$_\odot$ and metallicities of $z=0$, $z=10^{-4} z_\odot$, and $z= z_\odot$, where $z_\odot$ indicates solar metallicity. The solid-green curves show simulations that eventually explode, and the black curves show simulations that fail to explode.  The most striking feature in the FEC+ plot (right panel) is that there is a clear separation between failed and successful explosions for a value of FEC+ around 0.28--0.3.  This is yet another example showing that the FEC+ can be used as an explosion diagnostic. 

Although the definition of a successful explosion can vary, in this manuscript, we define a successful shock revival when the shock crosses 500 km. We will be using the same definition for the explosion time in  Section~\ref{sec:FEC_SiO_correlations}. All of the simulations that show a successful shock revival also have positive diagnostic energies $E^+_{\rm diag}$, where 
\begin{equation}
    \label{eq:diag_ene}
    E^+_{\rm diag} = \int_{e_{\rm bind} > 0} e_{\rm bind} dV - E_{\rm ov},
\end{equation}
where $e_{\rm bind}$ is the binding energy of the material and the integral is performed only over regions with positive binding energy \citep{Buras2006_2D_diag_ene,Bruenn2009_2D_3D_diag_ene}. Notice that we subtracted the binding energy (i.e. thermal energy minus gravitational energy) of the yet unshocked material ($E_{\rm ov}$) to have a more realistic estimate of the final explosion energy. The explosion energies at the end of the simulations of the 341 progenitors described above are shown in Figure~\ref{fig:diag_ene}. Given the relatively small radial domain of the simulations, the shock leaves the boundary before the diagnostic energy has plateaued, especially in the case of high-compactness progenitors, and therefore Figure~\ref{fig:diag_ene} represents a lower limit to the final explosion energy. 

In 1D simulations, the mass accretion is shut off once the shock is revived, and therefore neutrino heating drops to zero, even though a central black hole never forms. However, multi-dimensional simulations show that continued accretion after shock revival can be quite significant. This causes neutrino heating to be nonzero even after shock revival, potentially leading to larger explosion energies compared to 1D simulations \citep{Burrows2024_BH_formation_3D}. However, in the case of high-compactness progenitors, continued accretion causes an early collapse to a black hole, completely shutting off neutrino emission that, in some cases \citep{Powell2021_collapse_PISNe,Sykes2024_2D_fallback_SNe,Eggenberger2024_BHSNe_EOS}, can lead to a decrease of the final explosion energy. Overall, one can still see in Figure~\ref{fig:diag_ene} that the explosion energy in 1D+ simulations increases with the compactness of the progenitor, as also seen in multi-dimensional simulations \citep{Burrows2018_Physical_dependencies_SN}.

It is also instructive to analyze the ratio of non-exploding models compared to observational constraints. To do that, we assumed that stars below 11 $M_\odot$ (i.e. the lowest mass simulated for the $z=0$ and $z=10^{-4} z_\odot$ progenitors) would produce a successful explosion. Then, we weighted our simulations based on their initial ZAMS mass using a Salpeter initial mass function with exponent $-2.35$ \citep{Salpeter1955_IMF}, independently of metallicity. We calculated the fraction of failed SNe to be $f_{\rm fSNe} = 0.25-0.3$, depending on whether we assume the lowest mass of a star that explodes as a CCSN to be 8 $M_\odot$ or $9 M_\odot$, respectively (see Figure 6 in \citet{Boccioli2024_remnant}). This is consistent with the observed fraction of failed supernovae $f_{\rm fSNe} = 0.16^{+0.23}_{-0.12}$ \citep{Neustadt2021_failedSN_frac}.

In the remainder of this paper, we will use the FEC+ to diagnose why the Si/O interface can help facilitate explosions in many, but not all, successful explosions.

\begin{figure*}
\centering
\begin{subfigure}{0.33\textwidth}
    \includegraphics[width=\textwidth]{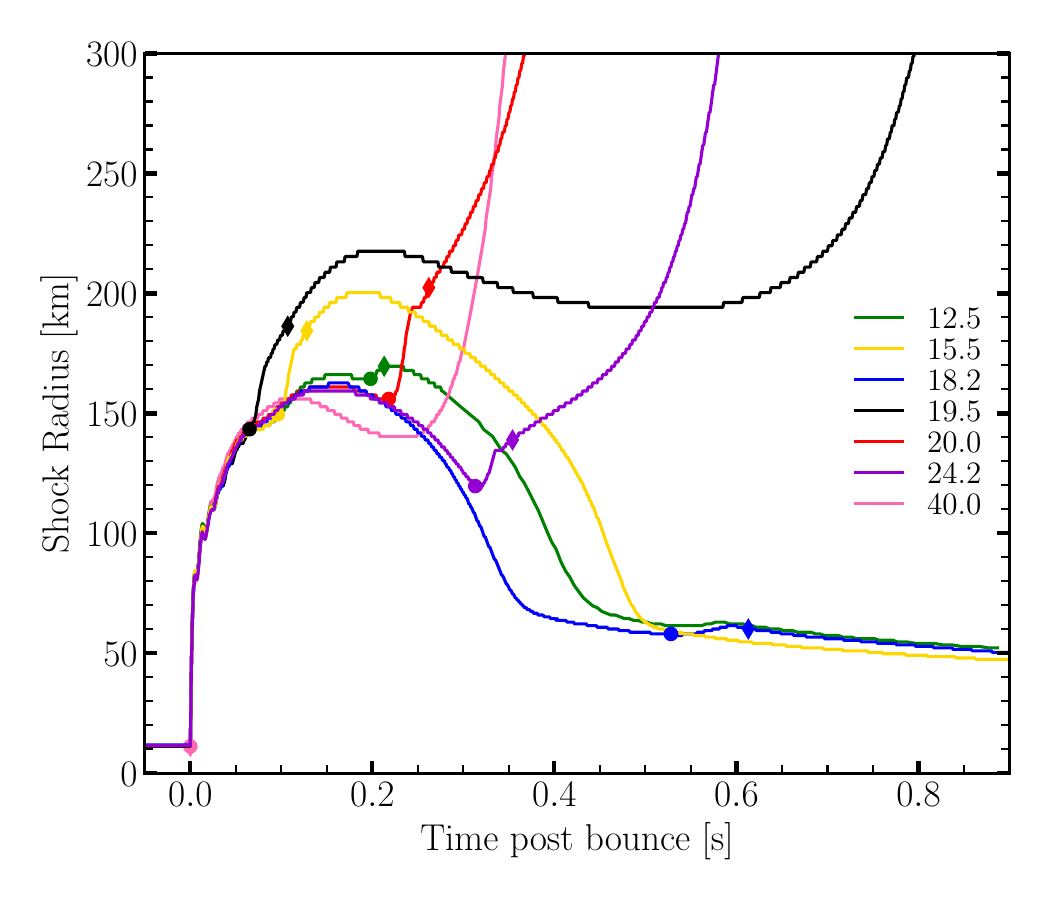}
    \caption{}
\end{subfigure}
\hfill
\begin{subfigure}{0.33\textwidth}
    \includegraphics[width=\textwidth]{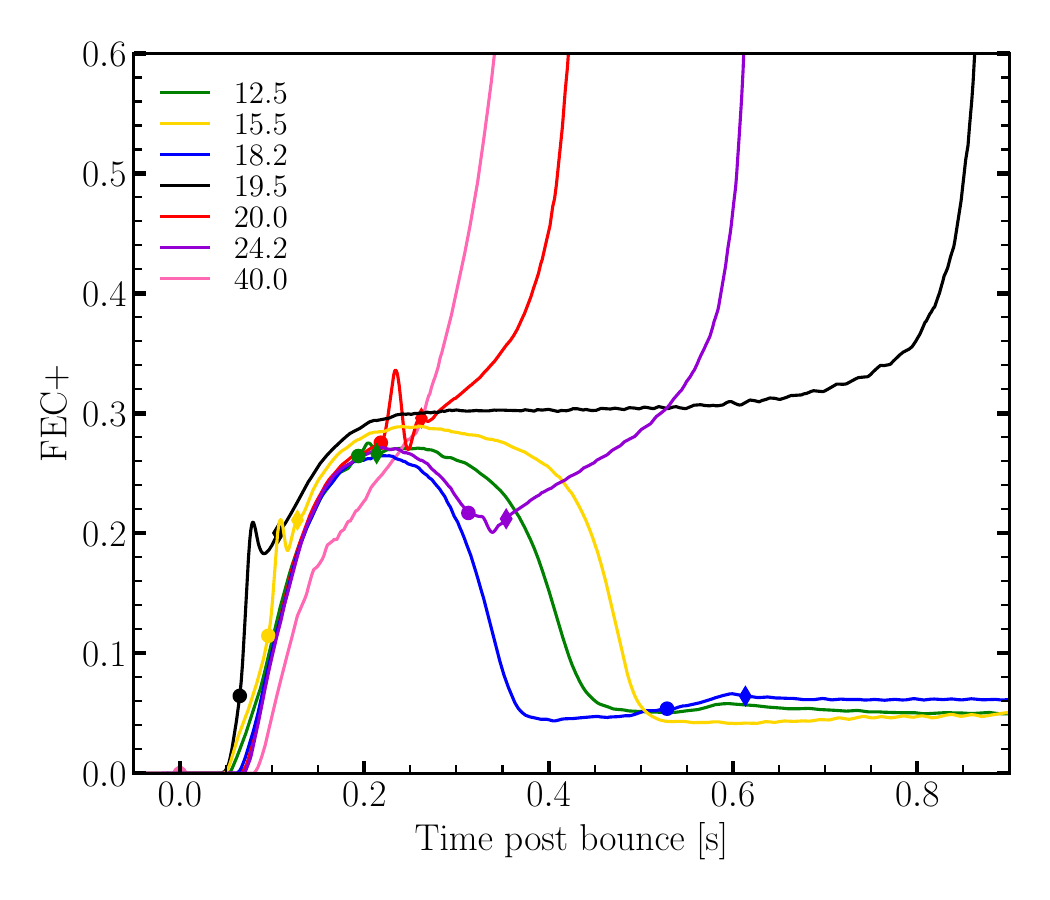}
    \caption{}
\end{subfigure}
\hfill
\begin{subfigure}{0.33\textwidth}
    \includegraphics[width=\textwidth]{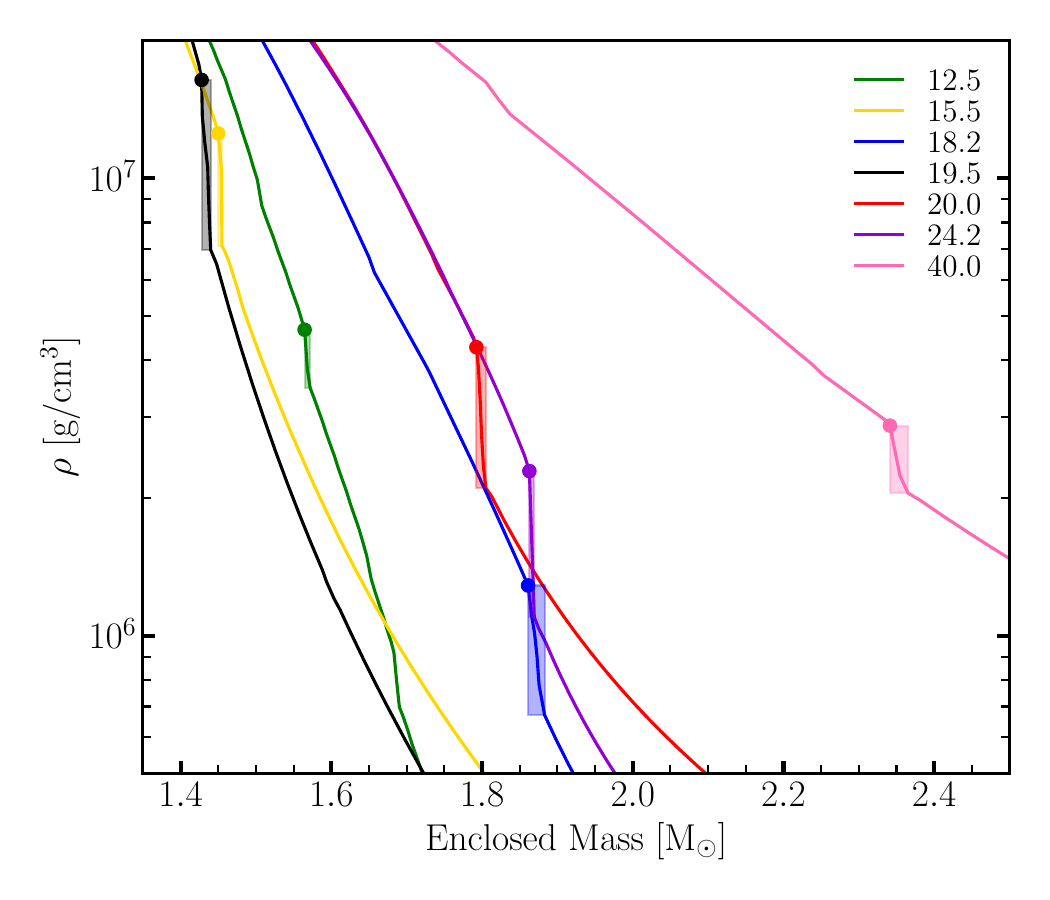}
    \caption{}
\end{subfigure}

\caption{This Figure shows the shock evolution, FEC+, and pre-collapse density structure of selected progenitors from the suite of 341 analyzed in this paper. Each progenitor is an example of a different explosive and non-explosive scenario, as discussed in the text (see Sec.~\ref{sec:FEC+_subset}). Moreover, we also show the 24.2 $M_\odot$ which does not fall in any of those scenarios, and it also has a very peculiar time evolution of the FEC+, which quickly dips similarly to what happens in failed SNe, before suddenly increasing after the accretion of the Si/O interface, which however does not seem to affect the FEC+. In the left and middle panels, circles and diamonds mark the times before and after the accretion of the Si/O interface, respectively. See Section~\ref{sec:change_alphaMLT} for a description of how these times are defined. The right panel shows the pre-SN density profiles for all of the progenitors, zoomed in near the Si/O interface. Circles mark the inner edge of the Si/O interface (which is used to define the time when the accretion stars $t_{\rm accr}^{\rm start}$), and the shaded regions show the extent of the interface. Notice that progenitors with lower compactness, have steeper density profiles.}
\label{fig:selected_progs}
\end{figure*}

\begin{figure}
\centering
\includegraphics[width=\columnwidth]{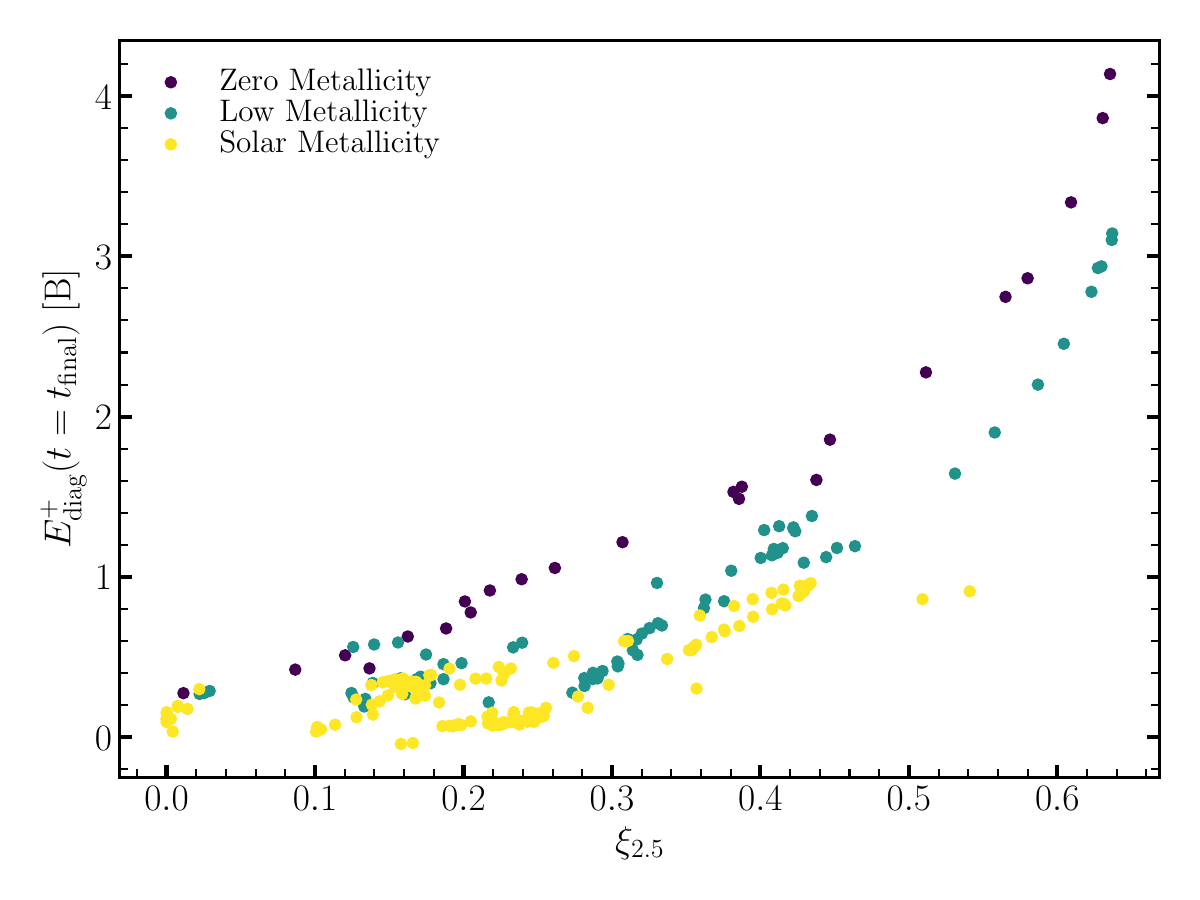}

\caption{Explosion energy calculated at the end of the simulations, as defined by Eq.~\eqref{eq:diag_ene}. This represents a lower limit of the explosion energy, since when the shock leaves the computational domain the diagnostic energy is still increasing. As expected the explosion energy increases with the progenitor compactness. The apparent trends of lower metallicity yielding large explosion energies for high-compactness progenitors, however, might be due to the aforementioned small computational domain preventing the diagnostic energy to plateau. Therefore, simulations with a larger radial domain (currently underway) are needed before drawing robust conclusions regarding the difference in explosion energy for progenitors of different metallicities.}
\label{fig:diag_ene}
\end{figure}

\subsection{Pre-explosion phase in light of the FEC+}
\label{sec:FEC+_subset}

For the present analysis, we analyze the post-bounce phase of the aforementioned 341 simulations. The role of the Si/O interface can be divided into three categories:  1) the accretion of the Si/O interface through the shock immediately triggers a vigorous explosion, 2) the accretion of the interface triggers outward movement of the shock, the shock stalls again, retreats a little, and then explodes, 3) the accretion of the interface triggers outward progression of the shock, which subsequently stalls again and does not explode. 

Generally, what determines which category a specific progenitor belongs to is how large the density drop at the Si/O interface is, and how late the interface is accreted. We consider a "large" interface one characterized by $\delta \rho_{\rm Si/O}^2/\rho_{\rm Si/O}^2 > 0.08$, consistent with the explodability criterion derived independently by \citet{Boccioli2023_explodability} and \citet{Wang2022_prog_study_ram_pressure}, as summarized in Section~\ref{sec:SiO_accr}. Moreover, in some high compactness ($\xi_{2.0}$) cases, explosions occur regardless of the accretion of the density drop. The compactness is defined as \citep{OConnor2011_explodability}:
\begin{equation}
    \label{eq:compactness}
    \xi_{\rm M} = \dfrac{M/M_\odot}{R(M)/1000\, {\rm km}},
\end{equation}
where $R(M)$ is the radial coordinate that encloses a mass $M$. 

In \citep{Boccioli2023_explodability}

Therefore, one can identify 6 different scenarios, shown by select simulations in Figure~\ref{fig:selected_progs}, that describe the post-bounce evolution of both exploding and non-exploding models:
\begin{enumerate}[label={\arabic*)}]
    \item A high-compactness progenitor, with $\xi_{2.0} \gtrsim 0.5$, achieves an explosion during the stalled phase independently of the presence of a large density drop in the pre-SN profile. This constitutes $\sim 26\%$ of all the analyzed simulations, including the 40 M$_\odot$ progenitor shown in Figure~\ref{fig:selected_progs}. 
    \item A progenitor with $\xi_{2.0} \lesssim 0.5$ achieves an explosion right after the accretion of a large density drop \textit{during} the stalled-shock phase. This constitutes $\sim 45\%$ of all the analyzed simulations, including the 20~M$_\odot$ progenitor shown in Figure~\ref{fig:selected_progs}. 
    \item A progenitor with $\xi_{2.0} \lesssim 0.5$ achieves an explosion significantly after the accretion of a large density drop \textit{before} the stalled-shock phase. This constitutes $\sim 2-3\%$ of all the analyzed simulations, including the 19.5 M$_\odot$ progenitor shown in Figure~\ref{fig:selected_progs}. 
    \item A progenitor with $\xi_{2.0} \lesssim 0.5$ fails to explode because it accretes a large (but not large enough) density drop \textit{before} the stalled-shock phase. This constitutes $\sim 4-5\%$ of all the analyzed simulations, including the 15.5 M$_\odot$ progenitor shown in Figure~\ref{fig:selected_progs}. Notice that these progenitors are very similar to the ones from the previous category but have smaller density drops (or the accretion occurs earlier).
    \item A progenitor with $\xi_{2.0} \lesssim 0.5$ fails to explode because it accretes a small density drop \textit{during} the stalled-shock phase. This constitutes $\sim 4-5\%$ of all the analyzed simulations, including the 12.5 M$_\odot$ progenitor shown in Figure~\ref{fig:selected_progs}. 
    \item A progenitor with $\xi_{2.0} \lesssim 0.5$ fails to explode because it accretes either a small or large density drop \textit{after} the stalled-shock phase. This constitutes $\sim 15\%$ of all the analyzed simulations, including the 18.2 M$_\odot$ progenitor shown in Figure~\ref{fig:selected_progs}. 
\end{enumerate}

The remaining $\sim 3-4\%$ of the analyzed simulations are outliers. For example, four progenitors with large compactness do not explode, whereas 10 progenitors with large Si/O interfaces accreted towards the beginning or the end of the stalled-shock phase do not explode. It is worth noting here that while these outliers do not conform to the Si/O interface condition for explosion, all simulations are consistent with the FEC+ explosion condition, i.e. the FEC+ never crosses the threshold. Perhaps the most interesting outlier is the 24.2 $M_\odot$ progenitor with $z=10^{-4} z_\odot$, which accretes a large density drop very late (and therefore is expected to fail) but instead leads to a successful explosion. We include this progenitor in Figure~\ref{fig:selected_progs} to highlight the unusual time evolution of the FEC+ and shock radius. Two more progenitors, i.e. the 23.4 and 23.2 $M_\odot$ with $z=10^{-4} z_\odot$ have a similar pre-SN structure and post-bounce evolution. However, these progenitors have high compactness ($\xi_{2.0} > 0.5$), and the Si/O interface is accreted a bit earlier. Therefore their shock radius and FEC+ time evolution are not quite as unusual as they are for the 24.2 $M_\odot$ progenitor. In other words, the FEC+ is not far from the threshold when the explosion sets in, whereas it clearly is for the 24.2 $M_\odot$ progenitor.

Since the FEC+ represents a clear threshold for explosion, and we know exactly which components contribute to its evolution, it is also a useful diagnostic to understand why the accretion of the Si/O interface initiates explosion when it does. In the two leftmost panels of Figure~\ref{fig:selected_progs}, we indicated the time before and after the accretion of the Si/O interface (hereafter $t_{\rm accr}^{\rm start}$ and $t_{\rm accr}^{\rm end}$, respectively) with circles and diamonds. We define $t_{\rm accr}^{\rm start}$ as the time when the inner side of the Si/O interface is accreted through the shock. A more detailed description of how we define $t_{\rm accr}^{\rm end}$ is given later in Section~\ref{sec:change_alphaMLT}. In the following, we analyze the FEC+ evolution for each of the above scenarios, showcased in Figure~\ref{fig:selected_progs}. 

The 40 M$_\odot$ progenitor accretes the Si/O interface significantly after the explosion has started. Therefore, the accretion of the Si/O interface does not play a role whatsoever in the explosion of this progenitor. This is confirmed by the smooth evolution of the FEC+, without any rapid increase caused by the accretion of the Si/O interface. What distinguishes this progenitor is its large pre-SN compactness $\xi_{2.0} = 0.75$ that causes the neutrino heating to be very large, which drives the FEC+ to go above the threshold and trigger the explosion. A more thorough analysis of the role of compactness in triggering the explosion is currently underway and beyond the scope of this paper.

The 20 M$_\odot$ progenitor has a large density drop at the Si/O interface, and therefore the shock revival is more pronounced than, for example, in the case of the 12.5 M$_\odot$ progenitor. This translates into a large increase of the FEC+ that is pushed above the threshold, indicating that an explosion occurs.


The 12.5 M$_\odot$ has a small density drop which is reflected in a small increase in the FEC+. However, this increase is not large enough to push the FEC+ above the threshold, and therefore the explosion eventually fails. 

The 18.2 M$_\odot$ has a large density drop, which is however accreted at $\sim$ 500 ms post bounce, after the stalled-shock phase of the shock has ended. As can be seen from the left panel of Figure~\ref{fig:selected_progs} the shock radius quickly recedes at very small radii after $\sim$ 250-300 ms. Therefore, even though a large interface is accreted at late times, the shock cannot be revived anymore. In multi-dimensional simulations, where not only neutrino-driven convection but also other instabilities and asymmetries can develop, the stalled-shock phase might be longer and/or more asymmetric, which could change the outcome of the simulation.

The 19.5 M$_\odot$ progenitor shows a unique behavior. A very large density drop is accreted early in the evolution. However, since the FEC+ is still small when the accretion occurs, it does not increase enough to go above the threshold. Therefore, the explosion is not triggered after the accretion of the interface, and in principle, one would expect the explosion to fail. However, since this early accretion causes the shock to stall at larger radii than any of the other progenitors, the FEC+ at around $150-200$ ms is the largest among the progenitors shown. Therefore, between 200 and 300 ms, the FEC+ is barely below the threshold. Eventually, at around 600-700 ms, the FEC+ finally goes above the threshold causing a (relatively) late explosion.

One can speculate about what the cause of the explosion is. For most progenitors, it is the accretion of the Si/O interface that directly launches the explosion by reducing the ram pressure on the shock. For this specific progenitor (as well as a handful of others), the explosion happens much later than the accretion. Nonetheless, the early accretion of the Si/O interface still plays a crucial role in causing the explosion. This can be seen by comparing the shock radius and FEC+ evolution for the 19.5~M$_\odot$ and the 15.5~M$_\odot$ progenitors. These have almost identical pre-SN density (and also compositional) structures, as seen in the third panel of Figure~\ref{fig:selected_progs}, with the important difference that the 15.5 M$_\odot$ progenitor has a smaller density drop at the Si/O interface, and also a Si shell that is 0.125 M$_\odot$ thicker, causing the interface to be accreted at a slightly later time. The smaller density drop causes the shock radius and the FEC+ to have a smaller increase, and therefore the FEC+ during the stalled-shock phase (i.e. at $\sim 200$ ms after bounce) is smaller for the 15.5 M$_\odot$ progenitor. This is the reason why the 15.5 M$_\odot$ progenitor fails whereas the 19.5 M$_\odot$ progenitor explodes at a much later time.

Another interesting feature of these two progenitors is that, since they have a very similar post-bounce evolution, the shock trajectory of the 15.5 M$_\odot$ progenitor would look pretty much exactly like the one for the 19.5 M$_\odot$ progenitor if one were to use a larger value of $\alpha_{\rm MLT}$. This shows that most CCSNe are in general very close to explosion, and small changes in the physics can lead to qualitatively drastically different outcomes. 

\begin{figure}
\centering
\includegraphics[width=\columnwidth]{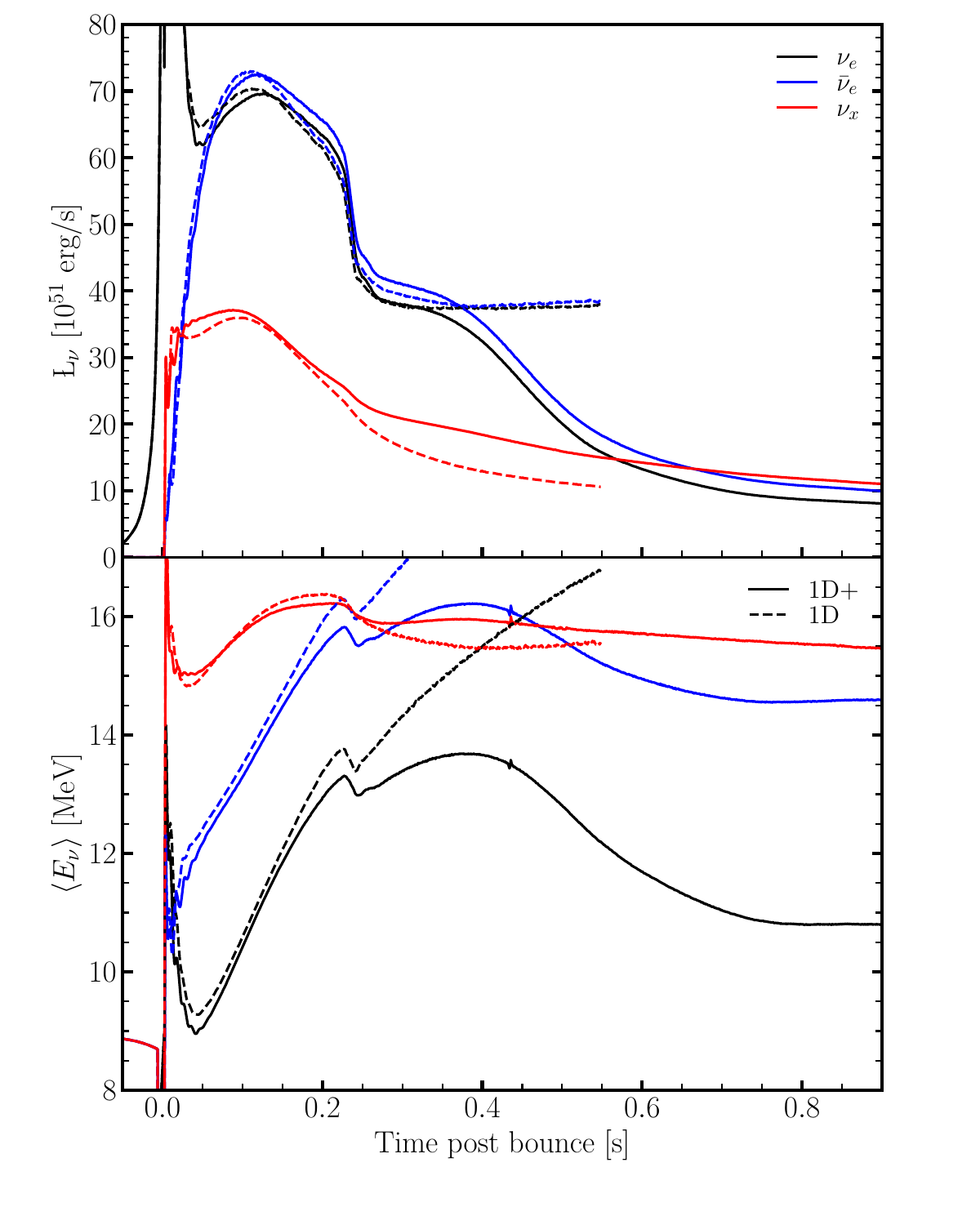}

\caption{The top panel shows the total net neutrino heating in the gain region for the set of progenitors discussed in Section~\ref{sec:FEC+_subset}. The bottom panel shows the ratio of the advection and heating timescales. Solid and dashed lines indicate simulations with and without STIR, respectively. Circles mark the times when the Si/O interface is accreted through the shock for the simulations with STIR. To avoid clutter, and since it occurs essentially at the same time, we did not mark the time of accretion of the Si/O interface for the simulations without STIR. The ratio of timescales, as expected, has some similarities to the FEC+ shown in the middle panel of Figure~\ref{fig:selected_progs}. However, as discussed in the text the FEC+ is a more robust, less ambiguous condition.}
\label{fig:Lnu_Enu_example}
\end{figure}

In order to better illustrate the explosion dynamics of 1D+ models, we show in Figure~\ref{fig:Lnu_Enu_example} the average energies and luminosities for both a 1D and a 1D+ simulation of a 20 $M_\odot$ progenitor from \citet{Sukhbold2016_explodability}, i.e. the same shown in Figure~\ref{fig:selected_progs}. The neutrino quantities for electron neutrinos and antineutrinos are quite similar in the 1D and 1D+ simulations since they are mostly determined by post-bounce mass accretion history, which is the same in 1D and 1D+ until the explosion sets in at around $\sim 220$~ms in the 1D+ model, causing mass accretion to stop and therefore neutrino energies and luminosities to decrease. In the case of heavy-lepton neutrinos, however, both average energies and luminosities are instead larger in the 1D+ model after the explosion sets in. This can be explained by the fact that heavy-lepton neutrinos are produced deeper in the PNS, and therefore are much less dependent on the accretion history of the progenitor \citep{OConnor2015}. 

The increase in the luminosity of heavy-lepton neutrinos at late times in 1D+ models can be attributed to PNS convection, which pushes the neutrinosphere at larger radii, increasing the area of emission, and therefore the luminosity. The reason for the larger neutrino energies in 1D+ simulations is instead more subtle. Convection in the PNS smooths out the density gradient, and therefore deep in the PNS, between $\sim 10^{13}$ and $10^{14}\ {\rm g/cm^3}$, the density gradient in the 1D case is shallower, whereas near the neutrinosphere, between $\sim 10^{12}$ and $10^{13}\ {\rm g/cm^3}$, it is steeper. The shallower gradient deep inside the PNS in the 1D simulation causes larger energy losses due to inelastic scattering, explaining the lower neutrino energies. This effect is subdominant for electron-flavor neutrinos since beta processes dominate the opacity.

\begin{figure}
\centering
\includegraphics[width=\columnwidth]{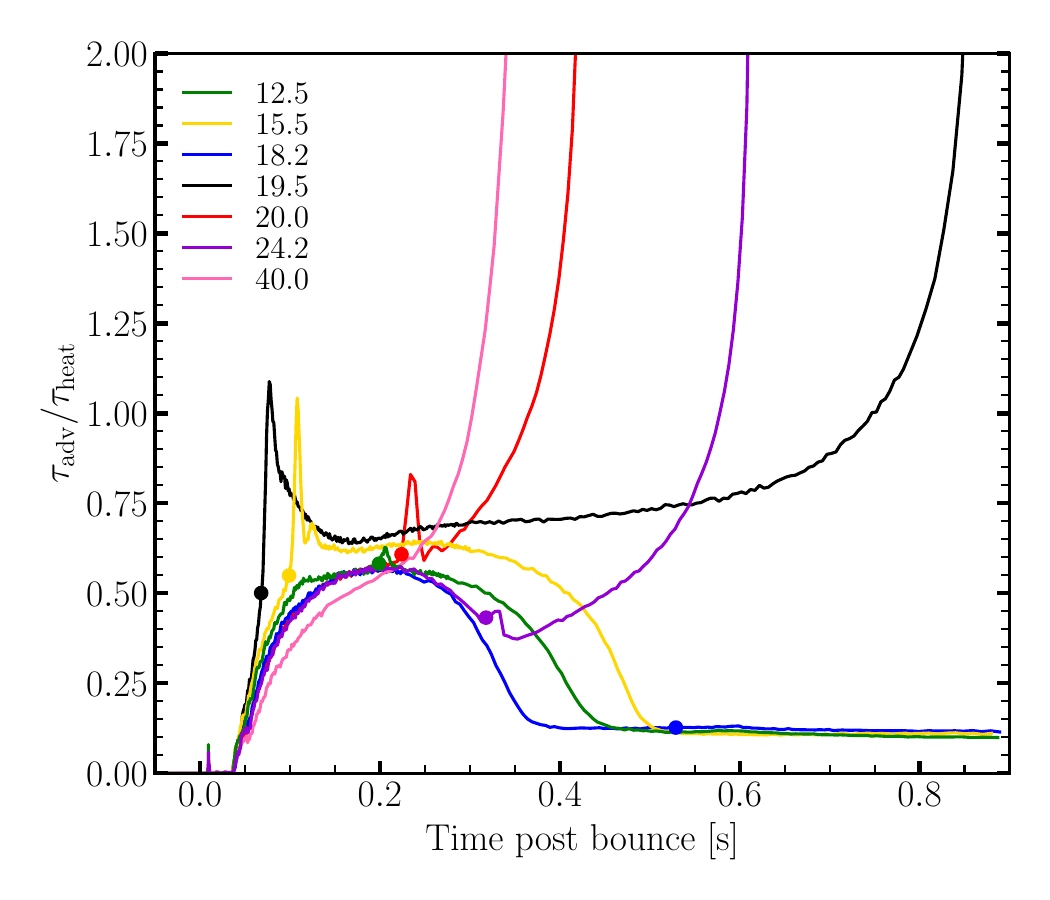}

\caption{The top panel shows the total net neutrino heating in the gain region for the set of progenitors discussed in Section~\ref{sec:FEC+_subset}. The bottom panel shows the ratio of the advection and heating timescales. Solid and dashed lines indicate simulations with and without STIR, respectively. Circles mark the times when the Si/O interface is accreted through the shock for the simulations with STIR. To avoid clutter, and since it occurs essentially at the same time, we did not mark the time of accretion of the Si/O interface for the simulations without STIR. The ratio of timescales, as expected, has some similarities to the FEC+ shown in the middle panel of Figure~\ref{fig:selected_progs}. However, as discussed in the text the FEC+ is a more robust, less ambiguous condition.}
\label{fig:timescales}
\end{figure}

A heuristic condition often used to determine whether an explosion sets in is based on the ratio of heating and advection timescales \citep{Fernandez2012_timescales,Buras2006_2D_diag_ene,Marek2009_SASI_diag_ene,Murphy2008_crit_lum_2D,Janka2001_conditions_shk_revival,Thompson2005_original_tau_crit}. A few different definitions of these two quantities are possible, and we adopt the following:
\begin{equation}
    \tau_{\rm adv} = \frac{M_{\rm gain}}{\dot{M}}~, \qquad \tau_{\rm heat} = \frac{E_{\rm gain}}{\dot{\mathcal{Q}}_\nu}
\end{equation}
where $E_{\rm gain}$, $M_{\rm gain}$, and $\dot{M}$ are the thermal energy, total mass, and average mass accretion rate in the gain region. It has been proposed that successful shock revival occurs when $\tau_{\rm adv} / \tau_{\rm heat} > 1$ \citep{Thompson2005_original_tau_crit}. Note that some groups \citet{Summa2018_rot3D_crit_lum} define $E_{\rm gain}$ as the total energy in the gain region (i.e. thermal plus kinetic plus gravitational) instead of the thermal energy only. However, in our case using the total energy results in the ratio $\tau_{\rm adv}/\tau_{\rm heat}$ to increase above 1 slightly earlier (and in some cases much earlier) than when shock revival occurs. This ambiguity is part of the reason why a more fundamental condition like the FEC+ should be preferred. We show the ratio of these two timescales in Figure~\ref{fig:timescales}, where one can also see the difference in adopting STIR (solid lines) compared to a simple 1D simulation (dashed lines).  Essentially, convection allows the shock to stall at a larger radius, causing neutrino heating to increase (i.e. $\tau_{\rm heat}$ to decrease) driving $\tau_{\rm adv}/\tau_{\rm heat}$ above 1. A more detailed study on how convection enables neutrino heating to increase is currently underway and is beyond the scope of this paper.

In the next section, we will investigate how changes in the strength of $\nu$-driven convection (i.e. changes in $\alpha_{\rm MLT}$) and in the density drop at the Si/O interface can change the shock and the evolution of the FEC+, leading to a successful or failed explosion. It is important to remark that also changes in the nuclear equation of state, neutrino opacities, asymmetries in the pre-SN progenitor, beyond standard model physics, collective neutrino oscillations, and more in general any physical input of the simulation, can lead to similar differences in the outcome of the supernova.

Since numerical simulations are involved, not only changes to the physics but also changes to the numerical setup and algorithms can lead to different outcomes. For example, as mentioned in the previous section, STIR has a weak dependence on resolution. Therefore, not only changes to $\alpha_{\rm MLT}$, but also changes to the numerical resolution can lead to an explosion for the higher resolution simulation and a failed supernova for the lower resolution simulation. This can be generalized to the case of multi-dimensional simulations, where numerical resolution is well-known to create different rates of turbulent dissipation \citep{Radice2016,Radice2018_turbulence,Couch2015_turbulence}, which can therefore affect the explodability.

To summarize, we qualitatively observed three types of explosions: (i) the ones triggered directly by the accretion of the Si/O interface (like the 20 M$_\odot$ in Figure~\ref{fig:selected_progs}), which causes a significant shock expansion and drives the FEC+ above the threshold, hence causing the explosion; (ii) the ones triggered indirectly by the accretion of the Si/O interface (like the 19.5 M$_\odot$ in Figure~\ref{fig:selected_progs}), where the accretion of this layer causes a significant shock expansion and pushes the FEC+ very close to the threshold until it eventually goes above and triggers the explosion; (iii) the ones triggered by a rapid increase of the FEC+, which occurs in high-compactness progenitors such as the 40~M$_\odot$ in Figure~\ref{fig:selected_progs}, not caused by accretions of any significant density drops, but simply a very strong neutrino heating. 

Most importantly, in all cases analyzed the explosion occurs after the FEC+ goes above $\sim 0.28\text{--}0.3$, despite the qualitative difference in how the FEC+ crosses this threshold, which shows the robustness and flexibility of the FEC+ as a diagnostic tool to determine the onset of explosion (but see the 24.2 $M_\odot$ progenitor as an exception to this statement).

\begin{figure}
\centering
\includegraphics[width=\columnwidth]{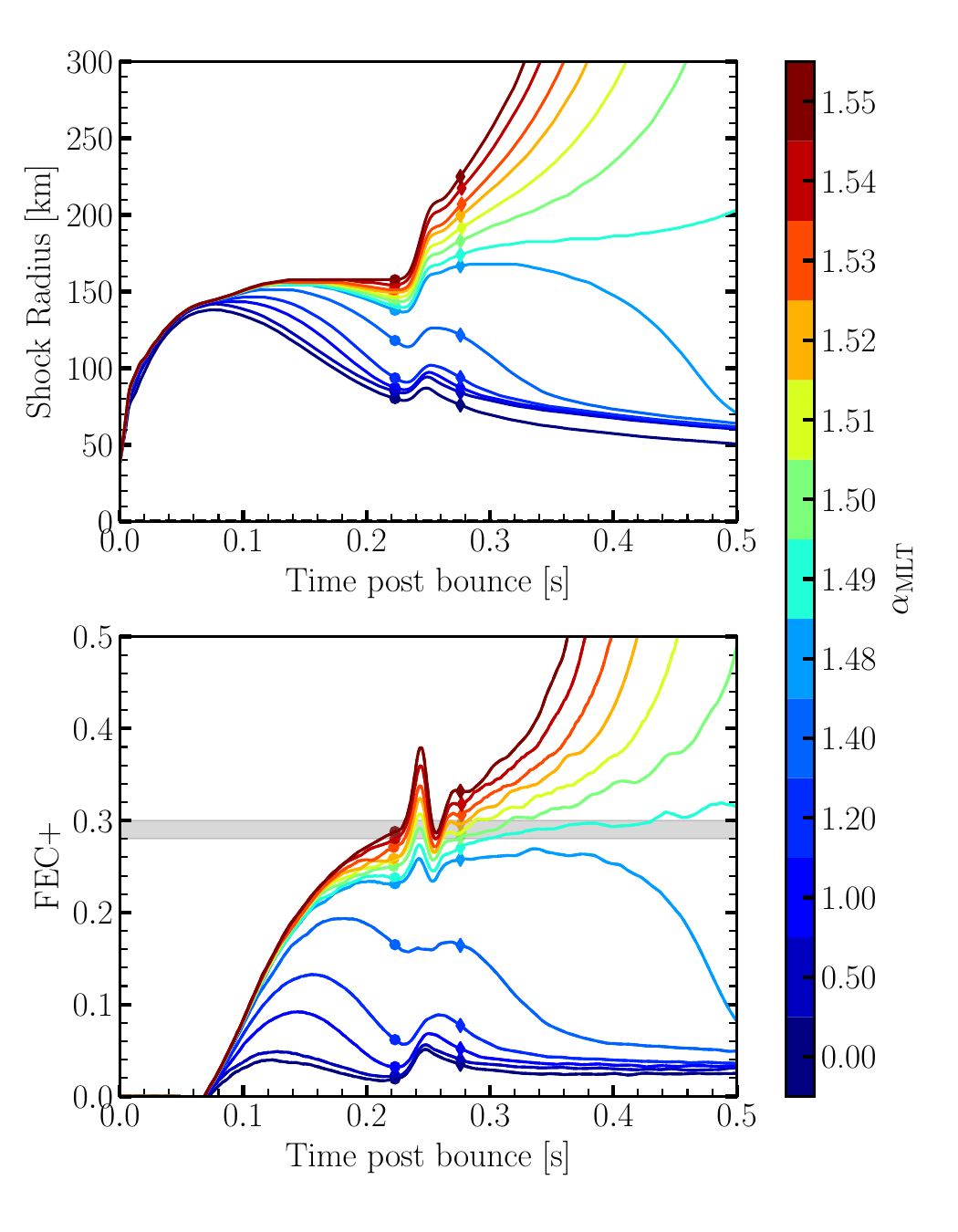}

\caption{The top (bottom) panel shows the evolution of the shock radius (FEC+) for a 20 $M_\odot$ MESA progenitor \citep{Farmer2016_preSN_properties} whose collapse was simulated using \texttt{GR1D+} for different values of $\alpha_{\rm MLT}$, shown in the colorbar on the right. Before calculating the value of the FEC+, each quantity in eq. \eqref{eq:FEC+} has been smoothed with a Savitzky-Golay filter in a 10 ms window. The filled dots show the time before the accretion of the Si/O interface, whereas the filled diamonds show the time after the accretion of the Si/O interface. All of the simulations with $\alpha_{\rm MLT} \geq 1.495$ successfully explode, and the accretion of the Si/O interface leads to a larger increase of the FEC+ for larger values of $\alpha_{\rm MLT}$. To avoid cluttering, we only show exploding simulations with $\alpha_{\rm MLT}$ spaced by 0.01, but a finer grid is shown in Figures \ref{fig:FEC_rshk_before_after_mesa} and \ref{fig:FEC_rshk_delta_mesa}.}
\label{fig:alphas_mesa}
\end{figure}

\section{Interplay between neutrino-driven convection and accretion of the Si/O interface}
\label{sec:interplay}
As shown in the previous section, the accretion of the Si/O interface can often lead to the explosion of a CCSN. In this section, we use the FEC+ to quantify in detail how the post-accretion expansion of the shock depends on both the strength of $\nu$-driven convection and the density drop at the Si/O interface. In addition to being an explosion condition, the FEC+ can also quantify a distance from explosion. Therefore, it is useful as a quantitative measure of how various physics affect the explosion outcome.  It should be highlighted that, in a more realistic 3D simulation, other multi-dimensional effects are also at play, and therefore the post-bounce phase is more complicated. However, the simplicity of 1D+ simulations enables a systematic study of the interplay between the accretion of the Si/O interface and parametrized $\nu$-driven convection.

\subsection{Sensitivity to the strength of neutrino-driven convection}
\label{sec:change_alphaMLT}
To analyze the interplay between (parameterized) neutrino-driven convection and the accretion of the Si/O interface, one can analyze how the FEC+ changes as a function of $\alpha_{\rm MLT}$. Therefore, we simulated the collapse and subsequent explosion of a 20 $M_\odot$ from \cite{Farmer2016_preSN_properties} for different values of $\alpha_{\rm MLT}$. This progenitor is particularly interesting as it is characterized by a large density drop at the Si/O interface. The shock evolution and the FEC+ as a function of time after bounce are shown in Figure~\ref{fig:alphas_mesa}. As in Figure~\ref{fig:all_progs}, one can see that there is a threshold for the FEC+ around 0.28--0.3 that separates explosions from failed SNe. As expected, simulations with small values of $\alpha_{\rm MLT}$ are further away from the threshold, and fail to explode. Simulations with large values of $\alpha_{\rm MLT}$ cross the critical threshold and lead to a successful explosion.  

\begin{figure}
    \includegraphics[width=\columnwidth]{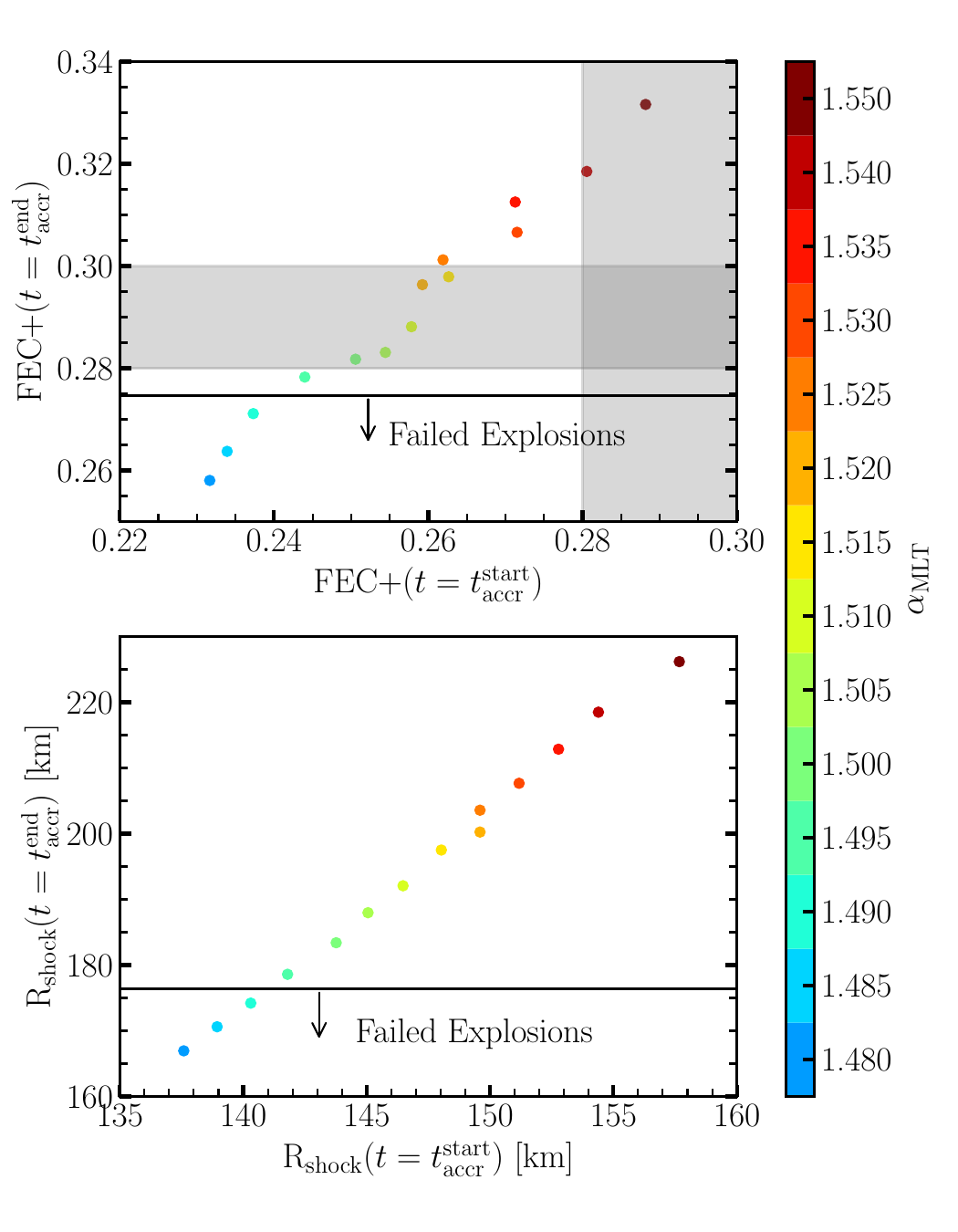}
\caption{The top panel shows the values of the FEC+ at $t=t_{\rm accr}^{\rm start}$ (x-axis) and at $t=t_{\rm accr}^{\rm end}$ (y-axis) the accretion of the Si/O interface through the shock. The times before and after accretion are shown as circles and diamonds in Figure~\ref{fig:alphas_mesa}. For a description of how exactly the times before and after accretion are defined, see the text. Only simulations with $\alpha_{\rm MLT} \geq 1.48$ are shown, since in other cases the FEC+ and shock radius do not show any significant "jumps". Simulations below the horizontal bar yield failed explosions, and indeed only when the FEC+ at $t=t_{\rm accr}^{\rm end}$ has crossed the threshold an explosion ensues. The bottom panel is the same as the top one but for shock radius instead of FEC+.}
\label{fig:FEC_rshk_before_after_mesa}
\end{figure}

For the accretion of the Si/O interface to be sufficient to launch the explosion, the FEC+ has to be already close to the threshold. This is shown in the top panel of Figure~\ref{fig:FEC_rshk_before_after_mesa}, where the x-axis is the value of the FEC+ at $t=t_{\rm accr}^{\rm start}$ (i.e. the circles in Figure~\ref{fig:alphas_mesa}) and the y-axis is the value of the FEC+ at $t=t_{\rm accr}^{\rm end}$ (i.e. the diamonds in Figure~\ref{fig:alphas_mesa}). Simulations with a value of FEC+ at $t=t_{\rm accr}^{\rm start}$ below $\sim 0.24$ do not explode, because once they accrete the interface the change in FEC+ is not enough to push it above the threshold. 

As can be seen in the bottom panel of Figure~\ref{fig:alphas_mesa}, the FEC+ peaks right after the accretion of the Si/O interface, before settling down to a more stable value. The reason behind this is that, as seen in the top panel, the accretion of the Si/O interface causes a sudden, extremely rapid expansion of the shock. In this transient phase, steady state is briefly disrupted and since the FEC+ is derived assuming a stalled shock solution, it cannot properly describe such rapid shock expansion, hence explaining the sudden peak in the FEC+.

To avoid the unexplained peak of the FEC+, we define the $t_{\rm accr}^{\rm end}$ to be the time for which the transient phase has passed \footnote{Specifically, we define $t_{\rm accr}^{\rm end}$ as the time $t>t_{\rm accr}^{\rm start}$ at which $\frac{{\rm d}^2 {\rm FEC+}}{ {\rm d} t^2 }$ has the second negative peak, which means a negative change in slope of the FEC+, indicating the end of the transient period, when the FEC+ reaches a (relatively) stable steady state. Notice that the first negative peak in $\frac{{\rm d}^2 {\rm FEC+}}{ {\rm d} t^2 }$ occurs in correspondence with the peak in the transient phase of the FEC+},  corresponding to the diamonds in Figures \ref{fig:alphas_mesa}.

All of the exploding simulations in Figure~\ref{fig:alphas_mesa} are characterized by a value of the FEC+ at $t=t_{\rm accr}^{\rm end}$ above 0.28--0.3. 

The bottom panel of Figure~\ref{fig:FEC_rshk_before_after_mesa} shows that the values of the FEC+ at $t=t_{\rm accr}^{\rm start}$ and $t=t_{\rm accr}^{\rm end}$ are related to the shock radii before and after the accretion, as one would expect. Moreover, both panels show a linear correlation between FEC+ and shock radius before and after accretion, and both quantities also positively correlate with $\alpha_{\rm MLT}$. The correlations for the FEC+ show a bit more scatter, although given the noise present in the FEC+ as a function of time (which has already been smoothed in Figure~\ref{fig:alphas_mesa}) it is not surprising.

\begin{figure}
    \includegraphics[width=\columnwidth]{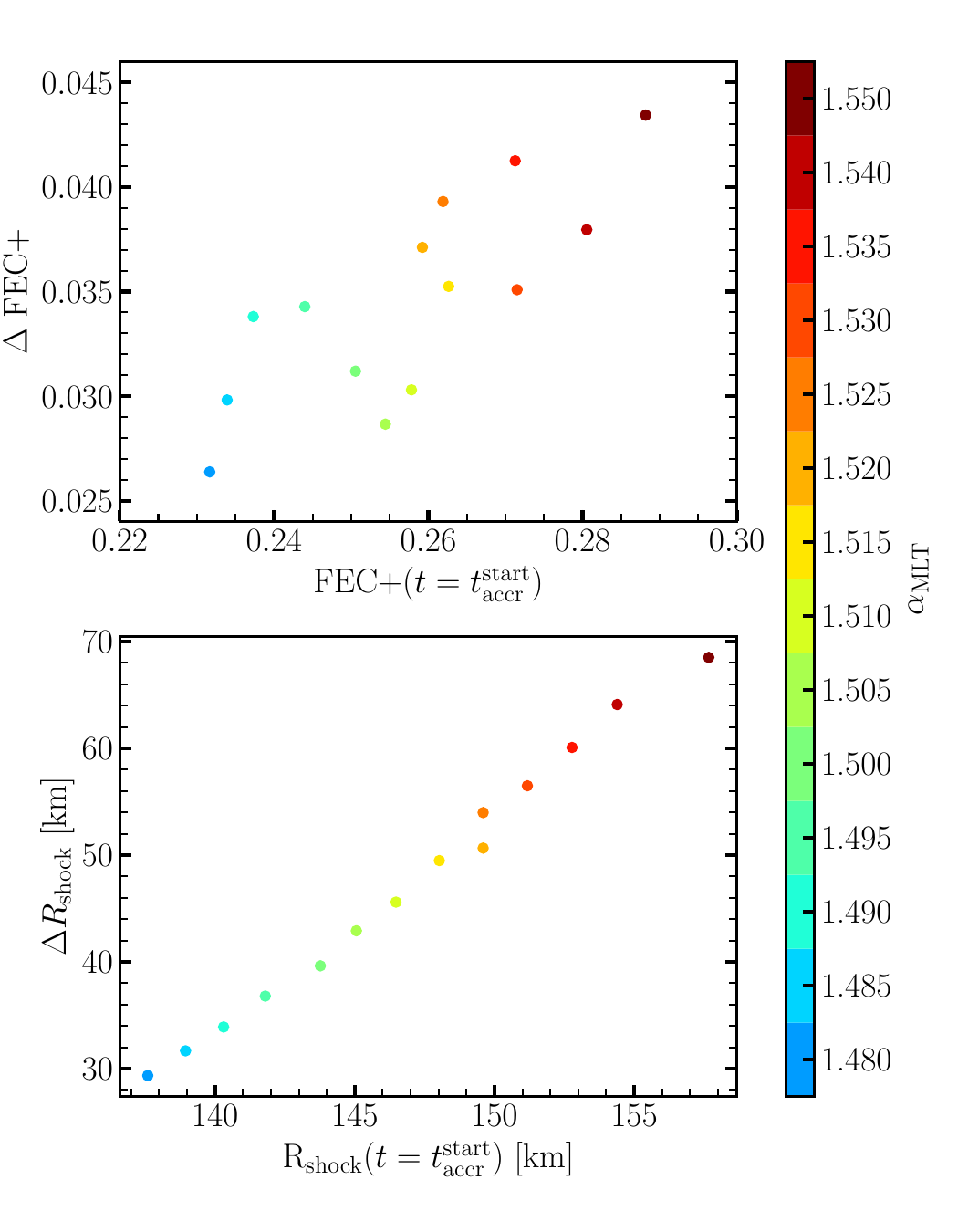}
\caption{The top panel shows on the y-axis the increase in FEC+ due to the accretion of the Si/O interface. This quantity, i.e. $\Delta {\rm FEC+}$, is simply the difference between the value of the FEC+ at $t=t_{\rm accr}^{\rm end}$ and at $t_{\rm accr}^{\rm start}$. The x-axis is the FEC+ at $t=t_{\rm accr}^{\rm start}$, the same as for Figure~\ref{fig:FEC_rshk_before_after_mesa}. The times before and after accretion are shown as circles and diamonds in Figure~\ref{fig:alphas_mesa}. For a description of how exactly the times before and after accretion are defined, see the text. Only simulations with $\alpha_{\rm MLT} \geq 1.48$ are shown, since in other cases the FEC+ and shock radius do not show any significant "jumps". The increase in FEC+ is proportional to $\alpha_{\rm MLT}$ and to the value of the FEC+ at $t_{\rm accr}^{\rm start}$, and for exploding progenitors with $\alpha_{\rm MLT}$ around 1.51 it is between 0.03 and 0.04, which is roughly $10\%$ of the FEC+ threshold. The bottom panel is the same as the top one but for shock radius instead of FEC+.}
\label{fig:FEC_rshk_delta_mesa}
\end{figure}

A similar correlation can be seen between $\Delta {\rm FEC+}$ (or $\Delta R_{\rm shock}$) and the value of FEC+ (or shock radius) at $t_{\rm accr}^{\rm start}$ (and also $\alpha_{\rm MLT}$), as shown in Figure~\ref{fig:FEC_rshk_delta_mesa}. The quantities $\Delta {\rm FEC+}$ and $\Delta R_{\rm shock}$ are defined as the differences between the values of FEC+ and shock radius after and before the accretion of the Si/O interface.
Larger $\alpha_{\rm MLT}$, i.e. stronger $\nu$-driven convection, leads to the shock stalling at a larger radius, causing the FEC+ to be larger, and eventually, the accretion of the Si/O interface causes a proportionally larger increase in FEC+ and shock radius.

Quantitatively, we observe that during the accretion, the Si/O interface increases the FEC+ by about 0.03--0.04 (top panel of Figure~\ref{fig:FEC_rshk_delta_mesa}) for $\alpha_{\rm MLT}$ around the best-value of 1.51. This increase is about 10\% of the FEC+ threshold, and it is about one-third of the overall effect of convection, which aids the explosion condition by about 25\text{--}30\% \citep{Gogilashvili2023_FEC+}.



To summarize, for a given density profile with a given Si/O interface, increasing $\alpha_{\rm MLT}$ will cause the shock to stall at larger radii, increasing the overall FEC+. When the Si/O interface is accreted, the shock expansion and the increase in FEC+ will also be larger. In particular, this increase is directly proportional to both $\alpha_{\rm MLT}$ and the radius of the shock (or value of FEC+) at $t_{\rm accr}^{\rm start}$. 

For a successful explosion to ensue, the FEC+ must be already close enough to the threshold (i.e. no more than $10\%$ below) when the Si/O interface is accreted. 

Another conclusion that can be drawn from the present analysis is that accurately simulating the effects of the convection is crucial since it contributes to determining the time evolution of the shock. This can significantly change the outcomes of the explosion because, as we have shown, the shock position before the accretion of the Si/O interface is strictly related to the effect of the Si/O interface accretion on the explosion condition.

In some 2D simulations, it has been shown that neutrino-driven convection is not properly captured, due to the inverse turbulent cascade caused by the imposed axisymmetry \citep{Couch2015_turbulence}. However, other studies have instead shown through detailed analysis of neutrino-driven convection that the turbulent dissipation in 2D simulations is compatible with the one found in 3D \citep{Murphy2013_turb_in_CCSNe,Mabanta2019_turbulence_in_CCSN}, and therefore the outcomes of the simulations should be the same. Moreover, 3D simulations can suffer from finite-resolution effects, and low resolution might artificially favor the explosion \citep{Abdikamalov2015_turb_SASI_3D,Radice2016}. Given the uncertainties and discrepancies among different works, more systematic studies and careful investigation of neutrino-driven convection effects are necessary to improve future multi-dimensional simulations.


\subsection{Sensitivity to the density drop at the Si/O interface}
\label{sec:smooth}

\begin{figure*}
\centering
\includegraphics[width=\textwidth]{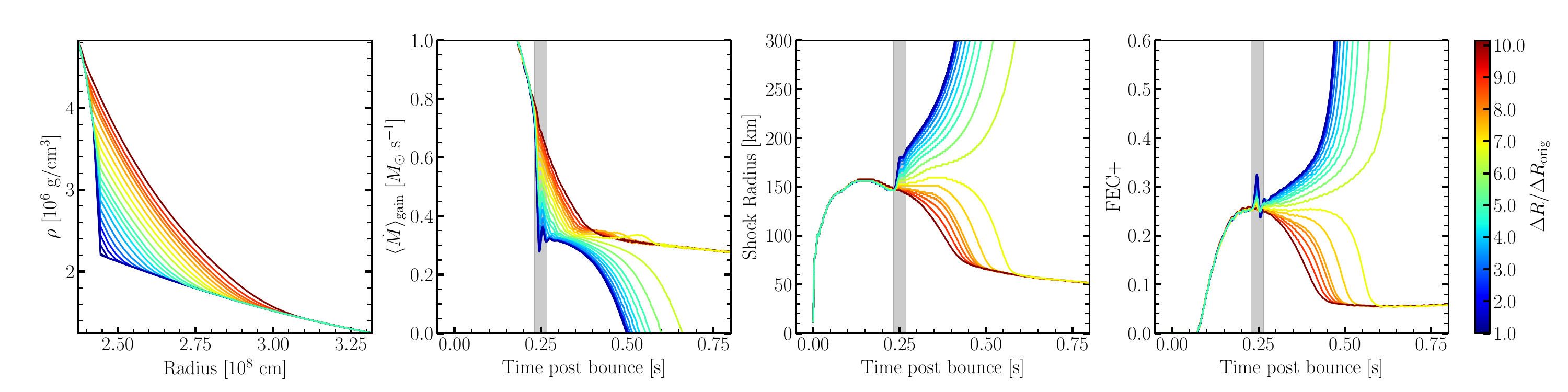}

\caption{This Figure refers to simulations of a 20 $M_\odot$ MESA progenitor \citep{Farmer2016_preSN_properties} whose density at the Si/O interface has been smoothed according to the procedure outlined in the text. Larger values of $\Delta R/ \Delta R_{\rm orig}$ correspond to a higher degree of smoothing. The first panel shows the pre-collapse density profile in the vicinity of the Si/O interface. The second panel shows the average mass accretion rate in the gain region as a function of time after bounce. The third panel shows the evolution of the shock radius, and the last panel shows the FEC+. The shaded region is drawn between $t_{\rm accr}^{\rm start}$ and $t_{\rm accr}^{\rm end}$, i.e. the time when the inner side of the Si/O interface is accreted and the time after the accretion and transient phase of the FEC+ has ended, as defined in Section~\ref{sec:change_alphaMLT}.}
\label{fig:smooth_mesa20}
\end{figure*}

\begin{figure*}
\centering
\includegraphics[width=\textwidth]{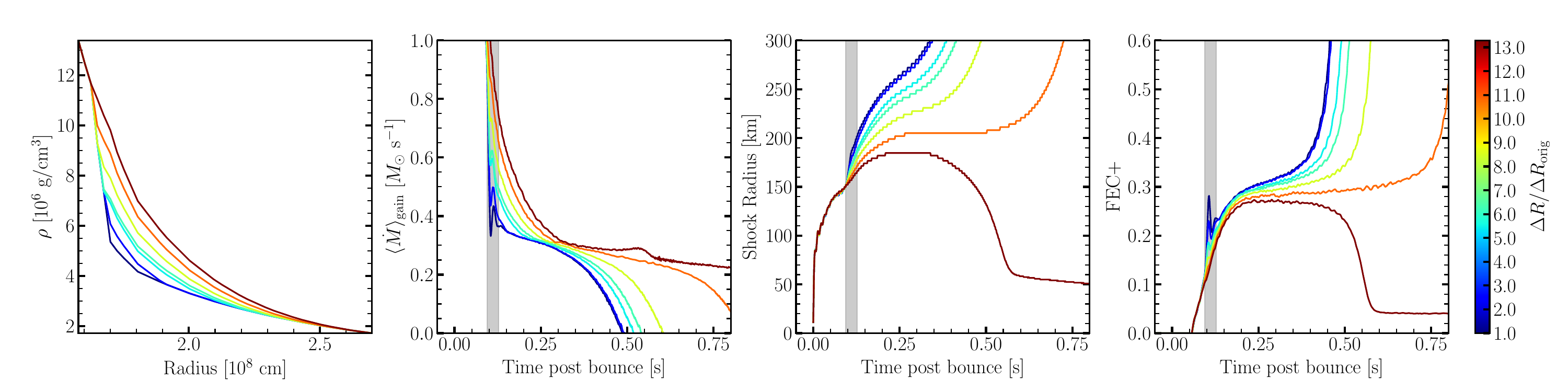}

\caption{Same as Figure~\ref{fig:smooth_mesa20}, but for a 21 $M_\odot$ KEPLER progenitor \citep{Sukhbold2016_explodability}}
\label{fig:smooth_S16_21}
\end{figure*}

Another way to observe the interplay between (parametrized) neutrino-driven convection and the accretion of the Si/O interface is to study how the explosion and the FEC+ change when only the Si/O interface changes, with everything else being the same. Therefore, we artificially smoothed the density drop at the Si/O interface of two different progenitors. The first two panels of Figures \ref{fig:smooth_mesa20} and \ref{fig:smooth_S16_21} show the smoothing of a $20 M_{\odot}$ MESA progenitor \citep{Farmer2016_preSN_properties} and a $21 M_{\odot}$ KEPLER progenitor \citep{Sukhbold2016_explodability}, respectively. 

The "degree of smoothing" shown in the colorbar of those figures is represented by the quantity $\Delta R/ \Delta R_{\rm orig}$, where $\Delta R$ is the radial extent of the Si/O interface, and $\Delta R_{\rm orig}$ refers to the original progenitor without any smoothing. Therefore, a more pronounced smoothing corresponds to a larger $\Delta R$ where the density profile falls off less steeply.

After smoothing the density profile, we calculated the new pressure by imposing hydrostatic equilibrium, whereas we did not change the electron fraction $Y_{\rm e}$, which only changes by $\lesssim 1\%$ anyway. Finally, with the new density, pressure, and electron fraction, we calculated the rest of the thermodynamic quantities by inverting the equation of state. We did not change the collapse velocities. Notice that, with this procedure, we are effectively adding mass outside the Si/O interface, since the density of the smoothed profiles is larger. However, this corresponds to a maximum of 0.02 $M_\odot$ and 0.025 $M_\odot$ for the $20 M_{\odot}$ MESA progenitor and the $21 M_{\odot}$ KEPLER progenitor, respectively. The artificially added mass is therefore small enough that it will not perturb the evolution in any way other than the change in the density jump at the Si/O interface.

One caveat that should be mentioned is that in principle this is not self-consistent since the Si/O interface corresponds to a change in composition. Therefore, one should actively change the composition at the Si/O interface to obtain a fully consistent thermodynamic profile, which would require a self-consistent simulation. However, this is quite challenging and would potentially modify the entire thermodynamic profile, and not only the Si/O interface. Therefore, we adopted this simpler approach of directly modifying the density profile, which is the most important quantity responsible for changing the ram pressure and, consequently, the effect of the accretion of the Si/O interface through the shock.

The results of the simulations for the 20 $M_\odot$ MESA progenitor from \cite{Farmer2016_preSN_properties} and for the 21 $M_\odot$ KEPLER progenitor from \cite{Sukhbold2016_explodability} are shown in Figures \ref{fig:smooth_mesa20} and \ref{fig:smooth_S16_21}, respectively. These two progenitors were chosen because of their large interfaces and the number of computational zones within the interface, allowing for a more accurate smoothing procedure. Moreover, the accretion of the Si/O interface occurs at $\sim 0.22$ s after bounce for the 20 $M_\odot$ MESA progenitor, and at $0.1$ s post bounce for the 21 $M_\odot$ KEPLER progenitor. This allows us to study how the time of accretion changes the impact of the interface on the explosion. 

The second panel of Figures~\ref{fig:smooth_mesa20}~\&~\ref{fig:smooth_S16_21} shows the average mass accretion rate in the gain region and, as expected, a smoother density profile at the Si/O interface corresponds to a shallower drop in mass accretion rate, which therefore prevents the shock from expanding. The third panels of those figures show the shock evolution for progenitors with different degrees of smoothing. The original progenitors explode and as the smoothing increases the shock expansion caused by the accretion of the Si/O interface becomes smaller, until eventually, it is not large enough to trigger an explosion. This can be seen by looking at the last panel, which shows that for the unmodified progenitor, the FEC+ goes above the threshold as a consequence of the accretion of the Si/O interface, resulting in an explosion.  As the smoothing increases, the increase is smaller and smaller until the FEC+ is not pushed above the threshold, and therefore the star does not explode.

The 20 $M_\odot$ MESA progenitor and the 21 $M_\odot$ KEPLER progenitor show the same general behavior concerning smoothing, and the threshold of the FEC+ is at around 0.28--0.3, consistent with the one found in the simulations discussed in sections \ref{sec:FEC_many_progs} and \ref{sec:change_alphaMLT}. Moreover, the 21 $M_\odot$ KEPLER progenitor shows an early accretion of the Si/O interface, much like the 19.5 $M_\odot$ progenitor shown in Figure~\ref{fig:selected_progs}, the difference being that the Si/O interface for the 21 $M_\odot$ progenitor has a larger $\delta\rho/\rho$, and therefore explodes early. However, as the degree of smoothing increases, one can see that for $\Delta R/ \Delta R_{\rm orig} \approx 10 $ (i.e. the orange curve in Figure~\ref{fig:smooth_S16_21}) the 21 $M_\odot$ progenitor has very similar behavior to the 19.5 $M_\odot$ progenitor. An early accretion of a relatively large Si/O interface leads to the shock stalling at a large radius, and the FEC+ being pushed just below threshold. Eventually, the FEC+ slowly increases and the star explodes a few hundred milliseconds later. If one increases the smoothing further, the star does not explode. 

This confirms that the explosion scenario (ii) in Section~\ref{sec:FEC_many_progs}, i.e. where the explosion is indirectly triggered by an early accretion of a large Si/O interface followed by a late explosion, is indeed an "edge case". The accretion of the Si/O interface needs to be quite early, and the density drop should not be too small, like in the case of the 15.5 $M_\odot$ progenitor in Figure~\ref{fig:selected_progs}. It should also not be too large, like in the case of the 21 $M_\odot$ progenitor studied here, which eventually behaves like the 19.5 $M_\odot$ progenitor in Figure~\ref{fig:selected_progs} only when the density drop is artificially smoothed.

This study shows how important it is to properly calculate the Si/O interface. This is no easy feat, given the complexity of stellar evolution calculations. For one these shells may or may not be a consequence of how 1D stellar evolution models handle the boundaries between stable layers and convectively unstable layers, which are inherently 3D, and even in simple 1D calculations, shell mergers might occur depending on how the convective boundaries evolve. In addition, it is well known that reduced burning networks, which decrease the computational cost and facilitate convergence, can be quite unreliable at the late stages of post-main sequence evolution of massive stars \citep{Farmer2016_preSN_properties,Renzo2024_small_networks}.

\section{Correlations between Si/O interface and FEC+}
\label{sec:FEC_SiO_correlations}
In Section~\ref{sec:change_alphaMLT} we showed how the change of the FEC+ induced by the accretion of the Si/O interface is proportional to the strength of neutrino-driven convection, and also to the value of the FEC+ itself right before the accretion. In this section, we quantify how the Si/O interface impacts the explosion by calculating the changes in the FEC+ caused by the accretion of the Si/O interface across the 341 progenitors analyzed in Section~\ref{sec:FEC_many_progs}. 

\begin{figure*}
\centering
\includegraphics[width=\textwidth]{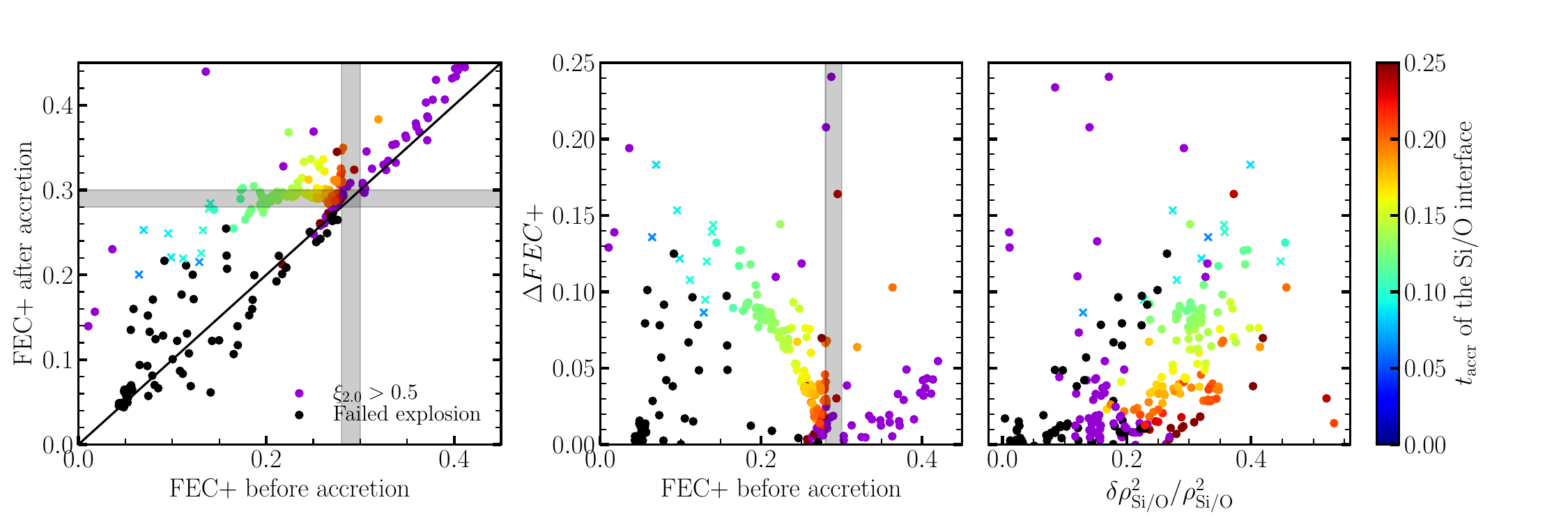}

\caption{The left panel shows the values of the FEC+ at $t=t_{\rm accr}^{\rm start}$ and at $t=t_{\rm accr}^{\rm end}$. The FEC+ threshold is shown as a shaded region between 0.28 and 0.3 on both the x and y-axis. The middle and right panels share the y-axis, which shows $\Delta {\rm FEC+}$, defined as the difference between the FEC+ at $t=t_{\rm accr}^{\rm end}$ and at $t=t_{\rm accr}^{\rm start}$. Finally, the right panel has $\delta \rho_{\rm Si/O}^2/\rho_{\rm Si/O}^2$ on the x-axis, where $\delta_\rho$ represents the density jump at the Si/O interface in the pre-SN progenitors, and $\rho_{\rm Si/O}$ is the density at the inner edge of the interface. The color of the dots highlights three scenarios: 1) the simulation fails to explode even after accreting the Si/O interface (black dots); 2) the simulation explodes even before the Si/O interface accretes (purple dots); 3) the Si/O interface triggers an explosion (dots and crosses color-coded depending on $t_{\rm accr}$). Crosses indicate progenitors with $t_{\rm accr} < 0.1$ s. }
\label{fig:FEC_before_after_rhoSiO}
\end{figure*}

First, one has to distinguish among the three major outcomes, displayed in Figure~\ref{fig:FEC_before_after_rhoSiO} with three different color schemes: 1) failed explosions shown as black dots; 2) high-compactness progenitors that usually explode before the accretion of the Si/O interface, shown as purple dots; 3) explosions caused by the accretion of the Si/O interface, shown as colored dots and crosses.

First, we focus our analysis on how the value of the FEC+ at $t=t_{\rm accr}^{\rm start}$ is related to the value of the FEC+ at $t=t_{\rm accr}^{\rm end}$ (left panel of Figure~\ref{fig:FEC_before_after_rhoSiO}). We define the time before (after) accretion as the time when the inner (outer) side of the Si/O interface, i.e. the left (right) edge of the rectangles in the right panel of Figure~\ref{fig:selected_progs} is accreted through the shock. This definition accurately captures the change in FEC+ for failed explosions (black dots) and high-compactness progenitors (purple dots). However, one needs to be more careful when the explosion occurs after the accretion of the Si/O interface (rainbow-colored dots). These cases are handled by defining the "time after accretion" as the time after which the transient phase has ended, as explained in Section~\ref{sec:change_alphaMLT}.

For the failed explosions (black dots) one expects scatter around the bisector. The reason is that when the Si/O interface is small (and therefore the explosion fails) it will not modify much the FEC+, and therefore the FEC+ before and after the Si/O interface is accreted through the shock should not be very different. The only exception is when the accretion happens during a rapid shock expansion (contraction), typically at early (late) times after bounce. By the time the outer edge of the Si/O interface is accreted, the FEC+ has therefore changed significantly. This explains the points above and below the bisector, for which the accretion occurs at early and late times, respectively.

For the high-compactness progenitors, shown as purple dots, one expects the FEC+ at $t=t_{\rm accr}^{\rm start}$ to be already above the threshold since, in most cases, the explosion sets in before the accretion of the Si/O interface. This is indeed confirmed by the fact that most of the purple dots are located at values of FEC+ at $t=t_{\rm accr}^{\rm start}$ larger than 0.3, indicating that the explosion has already set in. Note that there are three isolated cases in which accretion of the Si/O interface happens very early in the evolution while the shock is still expanding and has not yet reached the stalled shock phase. In these cases, while the early accretion of the Si/O interface changes the FEC+, the main change is due to the expansion of the shock.

Finally, the exploding progenitors with $\xi_{2.0} < 0.5$ are the remaining dots and crosses, color-coded based on the time when the accretion of the Si/O interface occurs. For these, the Si/O interface triggers an explosion, and the FEC+ quantifies why. Before the accretion of the Si/O interface, FEC+ is below the threshold, and after the accretion of the Si/O interface FEC+ is at or above the threshold. The progenitors shown as crosses are characterized by a very early accretion of the Si/O interface (see for example the 19.5 $M_\odot$ progenitor shown in Figure~\ref{fig:selected_progs}) and are therefore exceptions that will be discussed later in the section.
 
Similar conclusions can be drawn from the middle panel, in which $\Delta {\rm FEC+}$ (i.e. the difference between the FEC+ at $t=t_{\rm accr}^{\rm end}$ and at $t=t_{\rm accr}^{\rm start}$) is plotted against the value of the FEC+ at $t=t_{\rm accr}^{\rm start}$. The correlation between these two quantities is quite evident. The reason is that, if the accretion occurs early in the post-bounce phase, the shock is still expanding, and therefore the FEC+ is also increasing. Therefore, one expects larger $\Delta {\rm FEC+}$ since not only does the accretion increase the FEC+, but the secular shock expansion also contributes to the increase. This is confirmed by the fact that simulations with early accretions (i.e. $t_{\rm accr} \lesssim 0.2$ s) also exhibit larger $\Delta {\rm FEC+}$. As mentioned above, the reason is that if the accretion of the Si/O interface occurs early enough in the post-bounce phase, then the FEC+ increases not only because of the accretion but also because of the secular shock expansion. Notice that this discussion is not valid for failed explosions and high-compactness progenitors.

In the last panel, we show $\Delta {\rm FEC+}$ as a function of $\delta\rho_{\rm Si/O}^2/\rho_{\rm Si/O}^2$, where $\rho_{\rm Si/O}^2$ is the density on the inner side of the Si/O interface, and $\delta\rho_{\rm Si/O}$ is the density drop from the inner to the outer side of the interface. Examples of this are given in the rightmost panel of Figure~\ref{fig:selected_progs}. One would naively expect larger density drops to cause larger $\Delta {\rm FEC+}$, and indeed that trend is generally visible. However, it is also polluted by a large vertical scatter, which is due to early time accretions, for which not only the accretion of the Si/O interface, but also secular shock expansion contribute to $\Delta {\rm FEC+}$. This is confirmed by the fact that, given the same $\delta\rho_{\rm Si/O}^2/\rho_{\rm Si/O}^2$, simulations where the Si/O interface is accreted at early times lead to larger $\Delta {\rm FEC+}$. Therefore, if one takes into account the contribution to the FEC+ by secular shock expansion, one can find a much tighter dependence of $\Delta {\rm FEC+}$ on $\delta\rho_{\rm Si/O}^2/\rho_{\rm Si/O}^2$. By only selecting progenitors that have a relatively similar accretion time of the Si/O interface (i.e. similar colors in the rightmost panel of Figure~\ref{fig:FEC_before_after_rhoSiO}, one can indeed see that the vertical spread becomes much narrower.

Finally, there are a few exceptions worth discussing. The crosses in Figure ~\ref{fig:FEC_before_after_rhoSiO} are progenitors characterized by early accretion of the Si/O interface, i.e. $t_{\rm accr} < 0.1$ s. Their shock radius and FEC+ evolution are very similar to the 19.5 $M_\odot$ progenitor shown in Figure~\ref{fig:selected_progs}. In other words, they are part of scenario 3) defined in Section~\ref{sec:FEC_many_progs}, for which the early accretion of the Si/O interface does not push the FEC+ above the threshold, but instead pushes it very close to it, facilitating the occurrence of an explosion a few hundred milliseconds later. Therefore one expects the FEC+at $t=t_{\rm accr}^{\rm end}$ to be below the threshold, and the $\Delta {\rm FEC+}$ to be very large since the earlier the accretion, the faster the shock expansion, and therefore the larger the FEC+ increase. This is indeed confirmed by the left and middle panels in Figure~\ref{fig:FEC_before_after_rhoSiO}. Most of the simulations shown as rainbow-colored dots instead belong to scenario 2) defined in Section~\ref{sec:FEC_many_progs}. These are cases where the accretion of a large Si/O interface occurs during the stalled shock phase, pushes the FEC+ above the threshold, and therefore causes the explosion. As seen in the left panel of Figure~\ref{fig:FEC_before_after_rhoSiO}, for some of them the FEC+ at $t=t_{\rm accr}^{\rm end}$ is however below the threshold, and $t_{\rm accr} > 0.1$ s. These can be thought of as intermediate cases between scenarios 2) and 3). The accretion of the Si/O interface occurs at the very beginning of the stalled shock phase.

\begin{table*}
\centering 
\begin{tabular}{l|ccccccc}
\toprule
Condition & N$_{\rm sample}$ & Mean & Median & Q$_{10\%}$ & Q$_{25\%}$ & Q$_{75\%}$ & Q$_{90\%}$ \\
\midrule
All & 129 & 0.059 (20\%) & 0.053 (18\%) & 0.018 (6\%) & 0.034 (12\%) & 0.084 (29\%) & 0.096 (33\%) \\
$t_{\rm accr} > 0.12$ & 123 & 0.056 (19\%) & 0.048 (17\%) & 0.018 (6\%) & 0.033 (11\%) & 0.082 (28\%) & 0.090 (31\%) \\
$t_{\rm accr} > 0.15$ & 116 & 0.052 (18\%) & 0.044 (15\%) & 0.018 (6\%) & 0.032 (11\%) & 0.077 (27\%) & 0.088 (30\%) \\
$t_{\rm accr} > 0.18$ & 75 & 0.035 (12\%) & 0.035 (12\%) & 0.015 (5\%) & 0.022 (8\%) & 0.042 (15\%) & 0.061 (21\%) \\
$t_{\rm accr} > 0.2$ & 46 & 0.027 (9\%) & 0.027 (9\%) & 0.011 (4\%) & 0.018 (6\%) & 0.036 (12\%) & 0.039 (13\%) \\
FEC+ $> 0.2$ & 104 & 0.047 (16\%) & 0.039 (13\%) & 0.017 (6\%) & 0.029 (10\%) & 0.070 (24\%) & 0.084 (29\%) \\
FEC+ $> 0.22$ & 87 & 0.040 (14\%) & 0.037 (13\%) & 0.015 (5\%) & 0.024 (8\%) & 0.053 (18\%) & 0.070 (24\%) \\
FEC+ $> 0.24$ & 75 & 0.036 (12\%) & 0.035 (12\%) & 0.015 (5\%) & 0.022 (8\%) & 0.042 (15\%) & 0.064 (22\%) \\
FEC+ $> 0.26$ & 49 & 0.029 (10\%) & 0.029 (10\%) & 0.013 (4\%) & 0.018 (6\%) & 0.037 (13\%) & 0.041 (14\%) \\
\bottomrule
\end{tabular}

\caption{Statistical measures of $\Delta$FEC+ for different samples, which include all exploding progenitors with $\xi_{2.0} < 0.5$, and further restricted based on the condition described in the first column. The second column shows the size of the sample (i.e. how many simulations satisfy that condition), and the remaining columns are the mean, median, and 10\%, 25\%, 75\%, and 90\% quantiles. The percentage in parenthesis is calculated by dividing the number by the FEC+ threshold (i.e. 0.29), and it therefore shows how much the accretion of the Si/O interface contributes to the overall explosion condition. The first row includes all exploding progenitors with $\xi_{2.0} < 0.5$. The next four rows restrict the population further based on the accretion time of the Si/O interface. The last four rows restrict the population further based on the value of the FEC+ right before the accretion of the Si/O interface. The second to last row represents the most reliable population (see text). \label{tab:median_etc}}
\end{table*}

\subsection{Quantitative effect of Si/O interface accretion on the explosion condition}
In Section~\ref{sec:change_alphaMLT} we already compared the magnitude of $\Delta {\rm FEC+}$ with the overall explosion threshold of 0.28--0.3. However, that was done only for one progenitor, and we found that $\Delta {\rm FEC+}$ is about 10\% of the explosion condition. With the much larger sample of 341 progenitors, we can perform the same exercise and estimate the average impact that the Si/O interface has on the explosion condition. This is not a straightforward task since, as explained above, the secular shock expansion can sometimes overlap with the increase in FEC+ due to the Si/O interface accretion. Therefore, we exclude all of the high-compactness progenitors from the subsequent analysis, since as explained above the vast majority accreted the Si/O interface after the explosion has started. Then, we limit our sample to exploding progenitors that accrete the Si/O interface during the stalled-shock phase, when changes to the FEC+ can be attributed exclusively to the accretion of the interface, rather than to secular shock expansion. This can be achieved by excluding all progenitors for which the accretion of the interface occurs before a certain time or, alternatively, by excluding all progenitors for which the FEC+ at $t=t_{\rm accr}^{\rm start}$ is below a certain value. Depending on the chosen condition, the sample will be different. 

In Table~\ref{tab:median_etc} we show how different cuts on $t_{\rm accr}$ and the value of the FEC+ at $t=t_{\rm accr}^{\rm start}$ affect the sample size. Then, for the progenitors that satisfy those conditions, we calculate the median, mean, and selected quantiles of $\Delta {\rm FEC+}$. We find the conditions $t_{\rm accr} > 0.18$~s and FEC+~$> 0.24$ to lead to almost the same sample. We also consider them to be the best cuts since they include as many simulations as possible, without however including cases where significant secular shock expansion is occurring alongside the accretion of the Si/O interface. Notice that the condition that the FEC+ at $t=t_{\rm accr}^{\rm start}$ should be $> 0.24$ is the same found in Section~\ref{sec:change_alphaMLT} in order for the $20 M_\odot$ MESA progenitor to explode. For both samples, the median $\Delta {\rm FEC+}$ is $\sim 0.035$, which is in line with what was found for the $20 M_\odot$ MESA progenitor, and it is $\sim 10\%$ of the explosion condition, to be compared with the 25\text{--}30\% effect of convection \citep{Gogilashvili2023_FEC+}. The first and third quartiles are $\sim 0.023$ and  $\sim 0.046$, which correspond to $\sim 8\%$ and $\sim 15\%$ of the explosion condition, and more than 90\% of the simulations have $\Delta {\rm FEC+} > 0.015$, corresponding to $\sim 5\%$ of the explosion condition. This shows that the accretion of the Si/O interface has a crucial impact on the explosion of progenitors with compactness $\xi_{2.0} < 0.5$. This impact can be quantified to be between 5\% and 15\% of the overall explosion condition, or equivalently between $20 \%$ and $50 \%$ of the effect of convection.

\section{Conclusions}
\label{sec:conclusions}
In this paper, we analyzed the post-bounce phase of 341 1D+ simulations using the generalized Force Explosion Condition (FEC+). We found that when the FEC+ goes beyond the threshold, empirically found to be around 0.28--0.3, an explosion ensues. We identified three scenarios where an explosion is achieved: (1) High-compactness progenitors tend to always explode, often before accreting the Si/O interface; (2) lower-compactness progenitors, with $\xi_{2.0} \lesssim 0.5$, achieve an explosion if a large enough density drop ($\delta\rho_{\rm Si/O}^2/\rho_{\rm Si/O}^2$ > 0.08) is accreted through the shock during the stalled-shock phase; (3) a progenitor with $\xi_{2.0} \lesssim 0.5$ achieves an explosion significantly after the accretion of a large density drop before the stalled-shock phase. 

For failed explosions, we identified three scenarios: 4) a progenitor with $\xi_{2.0} \lesssim 0.5$ accretes a large density drop before the stalled-shock phase, although smaller than in scenario 3), and therefore not sufficient to cause an explosion; 5) a progenitor with $\xi_{2.0} \lesssim 0.5$ accretes a small density drop ($\delta\rho_{\rm Si/O}^2/\rho_{\rm Si/O}^2$ < 0.08) during the stalled-shock phase; 6) a progenitor with $\xi_{2.0} \lesssim 0.5$ accretes either a small or large density drop late in the stalled-shock phase when the shock cannot be revived anymore.

In all cases, the explosions ensue only when the FEC+ crosses the threshold of 0.28--0.3, showing that the FEC+ is a robust tool to describe the explosion condition.

We also studied how the FEC+ (and therefore the explosion) varies depending on the strength of $\nu$-driven convection (i.e. the mixing length parameter $\alpha_{\rm MLT}$). We again showed that the explosion occurs only when the FEC+ goes above the threshold of 0.28--0.3 empirically derived in Section~\ref{sec:FEC_many_progs}, which confirms the robustness of the FEC+ as an explosion condition. We showed that the value of the FEC+ at $t=t_{\rm accr}^{\rm start}$, as well as the change in FEC+ caused by the accretion of the Si/O interface, are positively correlated. They also positively correlate with $\alpha_{\rm MLT}$, the radius of the stalled shock, and the increase in shock radius caused by the accretion of the interface. This confirms the key role that convection has in aiding the explosion, which is extremely important since it has been shown that, for example, three-dimensional asymmetries in the pre-SN progenitor \citep{Muller2016_Oxburning,Yoshida2019_1D2D3D_Oburning,Yadav2020_ONe_convection,Couch2015_3D_final_stages,Fields2021_3D_burning_precollapse} can significantly change the strength of convection that develops in the post-bounce phase. Moreover, it is still unclear if and how numerical resolution \citep{Radice2016,Abdikamalov2015_turb_SASI_3D} or the dimensionality of the problem \citep{Couch2015_turbulence,Murphy2013_turb_in_CCSNe} can artificially alter the efficiency of convection.

Moreover, we show that the change in the FEC+ due to the accretion of the Si/O interface becomes progressively smaller by artificially smoothing the density drop at the Si/O interface. Eventually, for a large enough smoothing parameter, the FEC+ drops below the threshold causing the originally successful explosion to fail. Once again, we verified that the threshold in these simulations is also at around 0.28--0.3. We conclude that for progenitors with low to intermediate compactness ($\xi_{2.0} \lesssim 0.5$), the presence of a large density drop at the Si/O interface is crucial to determine whether or not an explosion occurs.

Finally, we analyzed how the change in FEC+ is related to the value of the FEC+ right before the accretion of the Si/O interface through the shock, and also to the density drop at the Si/O interface $\delta \rho_{\rm Si/O}^2/\rho_{\rm Si/O}^2$. We concluded that for the accretion of the Si/O interface to lead to successful explosions, the FEC+ should already be close to the explosion threshold, depending on how large the density drop at the Si/O interface is. We then quantified the effect of the Si/O interface accretion by analyzing the exploding progenitors with $\xi_{2.0} \lesssim 0.5$.

The analysis for progenitors accreting the interface early (i.e. before $\sim 0.18$~s after bounce) is complicated by the fact that the FEC+ changes not only because of the shock expansion caused by the accretion of the Si/O interface but also because of a secular shock expansion. By focusing only on progenitors where the secular shock expansion is negligible, we find that the accretion of the Si/O interface can contribute between 5 \% and 15\% to the overall explosion condition, i.e. between $20 \%$ and $50 \%$ of the overall effect of convection, which has been estimated to decrease the explosion condition by about 25\text{--}30 \% \citep{Gogilashvili2023_FEC+}.


The FEC+ is a powerful tool to analyze the post-bounce phase of the CCSN, and can be also applied to multi-dimensional simulations, where however more complex phenomena and geometries are at play, and therefore more careful analysis is required. These 1D+ simulations showed that the interplay between neutrino-driven convection and the accretion of the Si/O interface can be crucial for the explosion, and therefore future CCSN simulations should ensure that convection is properly resolved and carefully analyzed. 

Moreover, it is important to utilize (and also simulate!) pre-collapse models whose late stages of evolution have been carefully calculated. Since the density drop at the Si/O interface plays such a crucial role in the explosion, contributing to roughly 10 \% of the overall explosion condition, detailed stellar evolution models are required in order to accurately simulate the last few months of the life of massive stars, where shell Si-burning becomes important and can drastically modify the density discontinuity at the interface with the oxygen shell.

\section*{Acknowledgements}
The authors acknowledge the hosts and organizers of MICRA 2023 at ECT* for facilitating discussions on this project. L.B. is supported by the U.S. Department of Energy under Grant No. DE-SC0004658. L.B. would like to thank the N3AS center for their hospitality and support. E.O. is supported by the Swedish Research Council (Project No. 2020-00452). MG and JM acknowledge support from the Laboratory Directed Research and Development program, the Center for Space and Earth Sciences, and the center for Nonlinear Studies under project numbers 20240477CR-SES and 20220564ECR at Los Alamos National Laboratory (LANL). LANL is operated by Triad National Security, LLC, for the National Nuclear Security Administration of U.S. Department of Energy (Contract No. 89233218CNA000001). MG acknowledges support from the Danmarks Frie Forskningsfond (Project No. 8049-00038B, PI: I. Tamborra). This article is cleared for unlimited release, LA-UR-24-30985.

\section*{Data Availability}
The data is available upon reasonable request to the corresponding author.



\bibliographystyle{mnras}
\bibliography{References_Books,References_CNO,References_misc,References_EoS_neutrinos,References_Exp_Obs,References_Nucleosynthesis,References_SN,References_Stellar_Models} 






\bsp	
\label{lastpage}
\end{document}